\documentclass[a4paper,fleqn,usenatbib]{mnras}

\usepackage{newtxtext,newtxmath}
\usepackage[T1]{fontenc}
\usepackage{ae,aecompl}

\usepackage{graphicx}	% Including figure files
\usepackage{amsmath}	% Advanced maths commands
\usepackage{xcolor}           % Allows text colour for edits
\usepackage{xfrac}
\usepackage{gensymb}
\usepackage{pdflscape}
\usepackage{longtable}
\usepackage{subfig}

\newcommand{\Msun}{M$_{\odot}$}
\newcommand{\Mbh}{$M_{\rm BH}$}

\newcommand{\Mbulge}{$M_{\rm bulge}$}

\newcommand{\Lsun}{L$_\odot$}

\newcommand{\ml}{$M/L$}

\newcommand{\hst}{\emph{HST}}
\newcommand{\kms}{km~s$^{-1}$}

\mathchardef\mhyphen="2D

\title[Weighing central SMBH~in NGC~7469 using molecular and atomic gas dynamics]{Black hole mass measurement using ALMA observations of [CI] and CO emissions in the Seyfert 1 galaxy NGC~7469} 

\author[Dieu D.\ Nguyen et al.]{
Dieu D.\ Nguyen,$^{1,2}$\thanks{E-mail: dieu.nguyenduc@phenikaa-uni.edu.vn}
Takuma Izumi,$^{3,4}$
Sabine Thater,$^{5}$
Masatoshi Imanishi,$^{3,4}$
\newauthor{
Taiki Kawamuro,$^{3,6}$
Shunsuke Baba,$^{3}$ 
Suzuka Nakano,$^{3}$ 
Jean L.\ Turner,$^{7}$ 
Kotaro Kohno,$^{8,9}$ }
\newauthor{
Satoki Matsushita,$^{10}$
Sergio Mart\'in,$^{11,12}$
David S.\ Meier,$^{13}$
Phuong M.\ Nguyen,$^{14}$} and 
 \newauthor{
Lam T.\ Nguyen$^{15}$ }
\\
$^{1}$Faculty of Fundamental Sciences, PHENIKAA University, Yen Nghia, Hadong, Hanoi 12116, Vietnam\\
$^{2}$PHENIKAA Institute for Advanced Study (PIAS), Phenikaa University, Yen Nghia, Hadong, Hanoi 12116, Vietnam\\
$^{3}$National Astronomical Observatory of Japan (NAOJ), National Institute of Natural Sciences (NINS), 2-21-1 Osawa, Mitaka, Tokyo 181-8588, Japan\\
$^{4}$Department of Astronomical Science, The Graduate University for Advanced Studies (SOKENDAI), 2-21-1 Osawa, Mitaka, Tokyo 181-8588, Japan\\
$^{5}$Department of Astrophysics, University of Vienna, T\"urkenschanzstrasse 17, 1180 Wien, Austria\\
$^{6}$Nu\'{c}leo de Astronom\'{i}a de la Facultad de Ingenier\'{i}a, Universidad Diego Portales, Av. Ej\'{e}ercito Libertador 441, Santiago, Chile\\
$^{7}$Department of Physics and Astronomy, University of California at Los Angeles (UCLA), Los Angeles, CA 90095-1547, USA\\
$^{8}$Institute of Astronomy, Graduate School of Science, The University of Tokyo, 2-21-1 Osawa, Mitaka, Tokyo 181-0015, Japan\\
$^{9}$Research centre for the Early Universe, Graduate School of Science, The University of Tokyo, 7-3-1 Hongo, Bunkyo, Tokyo 113-0033, Japan\\
$^{10}$Academia Sinica, Institute of Astronomy $\&$ Astrophysics (ASIAA), P.O. Box 23-141, Taipei 10617, Taiwan\\
$^{11}$European Southern Observatory, Alonso de C\'ordova, 3107, Vitacura, Santiago 763-0355, Chile\\
$^{12}$Joint ALMA Observatory, Alonso de C\'ordova, 3107, Vitacura, Santiago 763-0355, Chile\\
$^{13}$Department of Physics, New Mexico Institute of Mining and Technology, Socorro, NM 87801, USA\\
$^{14}$Department of Physics, Quy Nhon University, 170 An Duong Vuong, Quy Nhon, Vietnam\\
$^{15}$LESIA, Observatoire de Paris, Universit\'e PSL, CNRS, Sorbonne Universit\'e, Univ. Paris Diderot, Sorbonne Paris Cit\'e, 5 place Jules Janssen,\\
 92195 Meudon, France
}
\date{Accepted 2021 April 7; Revised 2021 February 28; Received 2021 January 28; in original form 2021 January 28}
\pubyear{2021}

\begin{document}
\label{firstpage}
\pagerange{\pageref{firstpage}--\pageref{lastpage}}
\maketitle

\begin{abstract}
\normalsize 
\noindent  \normalsize We present a supermassive black hole (SMBH) mass measurement in the Seyfert 1 galaxy NGC~7469 using Atacama Large Millimeter/submillimeter Array (ALMA) observations of the atomic-${\rm [CI]}$(1-0) and molecular-$^{12}$CO(1-0) emission lines at the spatial resolution of $\approx0\farcs3$ (or $\approx$ 100 pc). These emissions reveal that NGC~7469 hosts a circumnuclear gas disc (CND) with a ring-like structure and a two-arm/bi-symmetric spiral pattern within it, surrounded by a starbursting ring. The CND has a relatively low $\sigma_{\rm gas}/V\approx0.35$ ($r\lesssim0\farcs5$) and $\approx0.19$ ($r>0\farcs5$), suggesting that the gas is dynamically settled and suitable for dynamically deriving the mass of its central source. As is expected from X-ray dominated region (XDR) effects that dramatically increase an atomic carbon abundance by dissociating CO molecules, we suggest that the atomic [CI](1-0) emission is a better probe of SMBH masses than CO emission in AGNs. Our dynamical model using the ${\rm [CI]}$(1-0) kinematics yields a $M_{\rm BH}=1.78^{+2.69}_{-1.10}\times10^7$~\Msun\ and $M/L_{\rm F547M}=2.25^{+0.40}_{-0.43}$ (\Msun/\Lsun). The model using the $^{12}$CO(1-0) kinematics also gives a consistent \Mbh\ with a larger uncertainty, up to an order of magnitude, i.e.\ $M_{\rm BH}=1.60^{+11.52}_{-1.45}\times10^7$~\Msun. This newly dynamical \Mbh\ is  $\approx$ 2 times higher than the mass determined from the reverberation mapped (RM) method using emissions arising in the unresolved broad-line region (BLR). Given this new $M_{\rm BH}$, we are able to constrain the specific RM dimensionless scaling factor of $f=7.2^{+4.2}_{-3.4}$ for the AGN BLR in NGC~7469. The gas within the unresolved BLR thus has a Keplerian virial velocity component and the inclination of $i\approx11.0^\circ$$_{-2.5}^{+2.2}$, confirming its face-on orientation in a Seyfert 1 AGN by assuming a geometrically thin BLR model. 
\end{abstract} 


%%%%%%%%%%%%%%%%%%%%%%%%%%%%%%%%%%%%%%%%%%%%%%%%%%

%%%%%%%%%%%%%%%%% BODY OF PAPER %%%%%%%%%%%%%%%%%%

%%%%%%%%%%%%%%%%%%%%%%%%%%%%%%%%%%%%%%%%%%
%%%%%%%%%%%%%%%%%%%%%%%%%%%%%%%%%%%%%%%%%%
\section{INTRODUCTION}\label{sec:intro} 

Supermassive black holes (SMBHs) appear to be ubiquitous at the centres of massive  galaxies. They have a strong influence on their bulge environments recorded in the scaling relations between the black hole mass (\Mbh) and the galaxy's macroscopic properties, e.g. stellar-velocity dispersion ($\sigma_\star$; \citealt{Gebhardt00, Ferrarese00}) and the bulge-stellar mass \cite[\Mbulge;][]{Kormendy95, Haring04, McConnell13}, despite the bulge extending well beyond the black hole's sphere of influence \citep[SOI;][]{Kormendy13}. These \Mbh--galaxy scaling relations suggest SMBHs and galaxies grow and evolve together through a series of accretion and merger events with feedback that regulates star formation \citep[e.g.][]{Schawinski07}.

The current data at the low-\Mbh\ regime ($<5\times10^7$~\Msun) hint a steeper \Mbh--\Mbulge\ slope for the lower-mass galaxies \citep[$M_\star<10^{10}$~\Msun;][]{Nguyen18, Nguyen19a, Nguyen20} despite a large scatter. One possible explanation is that black hole growth follows a different evolutionary track of the growth of bimodality-black hole seeds \citep[e.g.][]{Pacucci17, Pacucci18}. However, this black hole census remains incomplete, and typical scatter is about more than two orders of magnitude \citep[e.g.][]{Nguyen19a}. Additionally, another important but often unexplored point is the systematics in different methods that may bias \Mbh\ determinations from masers \citep[e.g.][]{Greene10, Kuo11}, stellar- and gas-dynamics \citep[e.g.][]{Krajnovic18, Thater19a}, reverberation mapping \citep[RM, e.g.][]{Peterson04}, velocity widths of broad optical emission lines \citep[e.g.][]{Baldassare15}. Often the different methods also do not give consistent results. Recently, there was a few direct comparisons but crucial to determine such bias, including (1) \Mbh\ in M87 inferred from stellar- \citep{Gebhardt11} versus ionized-gas kinematics \citep{Walsh13} using the observations from the Event Horizon Telescope \citep{EHT1, EHT4, EHT6} and (2) \Mbh\ in NGC~404 inferred from stellar- versus molecular-gas kinematics \citep{Davis20}. Any confident interpretation of low-mass black hole growth will require both larger numbers of \Mbh\ measurements and greater measurement precision.

The development of the cold-gas-dynamical method relies on high-spatial-resolution interferometric data at mm/(sub)mm wavelengths observed with Atacama Large Millimeter/submillimeter Array (ALMA), which has introduced a new powerful method to determine central $M_{\rm BH}$ in various types of galaxy morphologies and masses \citep{Davis14, Onishi15}. Most of these works utilised $^{12}$CO(2-1), which is bright and the best trade-off between sensitivity and resolution. Using the circumnuclear molecular gas disc's (CND, a massive gaseous disc with a few 100 pc in size) kinematics at the observational scale $\approx$ the black hole's SOI reduces the error associated with the galaxy-stellar-mass uncertainty \citep{Boizelle19, Boizelle20, North19}. Recent efforts of \citet{Davis20} and Nguyen et al. submitted have even pushed the possible of the method down to the lower-\Mbh\ regime of $\approx10^{5-6}$~\Msun. 

In this paper, we applied the cold-gas-dynamical method to estimate the central \Mbh\ in the Seyfert 1 active galactic nucleus (AGN) NGC~7469 using ALMA observations. NGC~7469 is one of the brightest type-1 AGNs, and therefore, its \Mbh\ was measured by various methods. However, the results range largely between $10^5$ and $10^8$~\Msun. The \Mbh\ of NGC~7469 was determined via the direct signatures of radio \citep[VLA 8.4 GHz;][]{Perez-Torres09} and X-ray \citep[{\it XMM-Newton} and {\it Swift};][]{Liu14} radiation, providing a $M_{\rm BH}=1.6^{+24.7}_{-1.5}\times10^7$~\Msun\ and $M_{\rm BH}=6_{-3}^{+27}\times10^5$~\Msun\ using the fundamental plane of \citet{Plotkin12} and \citet{Gultekin19}, respectively. Whereas, the RM method used the indirect signature from the broad line region (BLR) variability using H$\beta$ and \ion{He}{ii}$~\lambda$4686 emissions \citep{Peterson04, Peterson14, Wang14}, giving a $M_{\rm BH}=(1.0\pm0.1)\times10^7$~\Msun. Our simple but powerful method that uses cold gas dynamics will provide another direct constraint to its central black hole mass. Accurate \Mbh\ estimate in AGN, in turn, will shed light on the geometry/orientation of the ionizing photon-emitted BLR that is close to the central SMBH, which is unresolved with the current facilities but critical for the unified AGN model. 

As a follow-up to the ALMA emission-lines survey of the AGN in NGC~7469 \citep[][ hereafter I20]{Izumi20}, we utilised the bright, high signal-to-noise (S/N) and high-spatial-resolution data of the molecular-$^{12}$CO(1-0) ($\theta_{\rm beam}\approx0\farcs29$) and the atomic-${\rm [CI]}$(1-0) lines ($\theta_{\rm beam}\approx0\farcs31$) to perform cold-gas-dynamical models and weigh the central SMBH. We emphasise that the atomic-[CI](1-0) line is a potentially powerful tool to probe the nuclear gas dynamics in AGN. It is very bright at the CND-scale due to X-ray Dominated Region (XDR) processes around the AGN (i.e.\ a mass accreting SMBH) that include a highly efficient dissociation of CO molecules into C atoms \citep[e.g.][]{Maloney96, Meijerink05}. This process is recently confirmed in NGC 7469 as an elevated C/CO flux ratio and/or abundance ratio (I20). Hence, [CI](1-0) can potentially be a better probe of nuclear gas dynamics than $^{12}$CO(2-1) around AGNs and in determining \Mbh. The atomic-[CI](1-0) (rest frequency of $\approx492$ GHz) line has a higher rest-frame frequency than $^{12}$CO(2-1) (rest frequency of $\approx230$ GHz) or $^{12}$CO(1-0) (rest frequency of $\approx115$ GHz) and thus can be a suitable tracer to perform similar observations and dynamical modellings toward high-redshift AGNs.

The paper is organised into six Sections.  We present all crucial properties of NGC~7469 in Section~\ref{ngc7469} and \hst\ and ALMA observations in Section~\ref{data}. The mass modeling of NGC~7469 and the Kinematics Molecular Simulation \citep[KinMS;][]{Davis14} tool, used to estimate the central \Mbh, and the results are described in Section~\ref{kinms}. We then discuss the atomic-[CI](1-0) emission is probably a new gas-tracer (also better than the frequent use of low-J CO lines) to estimate \Mbh\ in the AGN and compare this new \Mbh\ with its RM-based estimates, as well as the usage this new \Mbh\ to constrain the inclination angle of the unresolved BLR in Section~\ref{discussion}. Finally, we conclude our findings in Section~\ref{conclusions}.

Throughout the article, all the maps are plotted with the orientation of north up and east to the left. We adopt an angular-size distance to NGC 7469 of $D_A=68.4\pm18.8$ Mpc, where the error is 1$\sigma$ scatter among 17 distances from NASA/IPAC Extragalactic Galaxies (NED\footnote{\url{https://ned.ipac.caltech.edu/}}), giving a physical scale of 330 pc arcsec$^{-1}$ in the standard flat Universe with $H_0\approx70$ \kms Mpc$^{-1}$, $\Omega_{\rm M}\approx0.3$, and $\Omega_{\rm \Lambda}\approx0.7$ (corresponding to a luminosity distance of $D_L=70.8$ Mpc, I20). The derived \Mbh\ scales linearly with the assumed distance, so any change to the distance will result in a compensated shift in \Mbh. We should emphasise that this is the lowest spatial-resolution limit at which we still can perform an accurate-\Mbh\ measurement using of the atomic-[CI](1-0) emission observed with ALMA. All quantities are quoted with the Galactic extinction correction to recover their intrinsic values assuming $A_V=0.184$ \citep{Schlafly11} and the interstellar extinction law of \citet{Cardelli89}.

%%%%%%%%%%%%%%%%%%%%%%%%%%%%%%%%%%%%%%%%%
%%%%%%%%%%%%%%%%%%%%%%%%%%%%%%%%%%%%%%%%%
\section{NGC~7469}\label{ngc7469}

NGC~7469 (Mrk~1514, Arp~298) is identified as a stellar-bar spiral, with a Hubble type (R$^{\prime}$)SAB(rs)a at near-infrared (NIR) wavelength \citep{Knapen00}. It is also classified as a luminous infrared galaxy (LIRG) based on its high infrared (IR) luminosity \citep[$L_{8-1000\;\mu m} = 10^{11.7}$~\Lsun;][]{Sanders03}.

Analysis from the molecular $^{12}$CO(1-0) and HCN(1-0) emissions observed with IRAM Plateau de Bure Interferometer (PdBI) interferometer \citep{Guilloteau92} yields a systemic velocity $V_{\rm LSR}=4925$~\kms, kinematic inclination $i=45^\circ$, and position angle PA~$=128^{\circ}$ \citep[][henceforth D04]{Davies04}.

Optical spectroscopy confirmed the nucleus of NGC 7469 hosts a luminous Seyfert 1 AGN traced by (i) strong emission at $K$-band 2.2 $\mu$m \citep{Genzel95, Lonsdale03, Imanishi04}, (ii) a core jet-like structure \citep[e.g.][]{Lonsdale03, Alberdi06} and ionized gas outflows \citep{Scott05, Blustin07}, and (iii) UV and X-ray variability \citep[e.g.][]{Kriss00, Nandra00, Petrucci04, Scott05}. This variability is also seen in optical broad Balmer lines \citep[FWHM$_{\rm  H\beta\;\lambda4861} = 4369$~\kms;][]{Bonatto90, Collier98, Peterson14}. \citet{Davies07} find a young stellar population ($\approx110$--$190$ Myr) in the NGC~7469's CND. The average star formation rate there is also high ($\langle SFR \rangle = 50-100$~\Msun\ yr$^{-1}$ kpc$^{-2}$), indicating a composite core of a type 1 AGN and a starburst ring distributed in the annulus of $1\farcs5-2\farcs5$ ($\approx500$--$833$ pc) at the central kpc region of this galaxy \citep{Soifer03, Diaz-Santos07}. \citet{Genzel95} argue that this nucleus region accounts for two-thirds of the galaxy bolometric luminosity \citep[$L_{\rm bol.} = 10^{45.3}$ erg s$^{-1}$;][]{Kaspi00}. This starburst ring was also clearly detected by ALMA high resolution ($<0\farcs6$) (sub)mm dust continuum data \citep{Izumi15, Imanishi16}.

\citet{Liu14} investigated the nuclear X-ray spectra observed by {\it XMM-Newton} and {\it Swift} of the NGC~7469 AGN, which originates from inverse Compton scattering excited by hot and compact corona near the SMBH. They measured hard/soft X-ray luminosities of $\log L_{\rm 2-10~keV}=43.170\pm0.009$ erg s$^{-1}$ and $\log L_{\rm 14-195~keV}=43.602^{+0.315}_{-1.184}$ erg s$^{-1}$. Additionally, \citet{Perez-Torres09} use VLA 8.4 GHz observations at $\approx0\farcs3$ resolution to estimate the radio emission of the nucleus of $\log L_{\rm 8.4~GHz}=36.959\pm0.009$ erg s$^{-1}$.  The radiations suggest that the AGN is shining at the Eddington ratio of $\approx0.3$ \citep{Petrucci04}.

The nuclear $^{12}$CO(1-0) gas forms a CND at the centre and a ring-like morphology located at $\approx$ $1\farcs$5--$2\farcs$5. Additionally, there is a bar or a pair of spiral arms between the ring and the CND with weak NIR emission \citep[D04;][]{Izumi15}. The total molecular gas mass inferred from $^{12}$CO(1-0) within the radius of $r\lesssim3\farcs5$ ($\approx1.2$ kpc) is $2.7\times10^9$~\Msun\ (D04, I20), while this mass of the entire galaxy is $M_{\rm H_2}\approx10^{10}$~\Msun\ \citep{Meixner90}. 

NGC~7469 has a bulge mass of $M_{\rm bulge}=(1.1\pm0.3)\times10^{11}$~\Msun\ (and $M_B=-20.9\pm0.2$ mag) based on the black hole-to-bulge mass relation in AGN \citep{Wandel02}, while its bulge-disc luminosity decomposition from \hst/$R$-band image gives $M_{R}=-22.08\pm0.75$ mag \citep{McLure00, McLure01}. Also, \citet{Onken04} used slit-spectroscopy to estimate its bulge/spheroid velocity dispersion of $\sigma_\star=152\pm16$~\kms\ using the stellar absorption lines of \ion{Ca}{ii} triplet (CaT) in the NIR regime excited by the AGN. 

We summarised these properties of NGC~7469 in Table \ref{ngc7469property}.

%%%%%%%%%%%%%% TABLE 1 NGC 7469 properties %%%%%%%%%%%%%
\begin{table}
\caption{Properties of NGC~7469}
\begin{tabular}{lcr}
\hline\hline   
Parameter (Unit)& Value & References\\
 (1)      &  (2)  & (3) \\
\hline
Morphology                                       &(R$^\prime$)SAB(rs)a&(1)\\
Nuclear activity                                 & Seyfert 1          &(2)\\
R.A. (ICRS)                                      &$23^{\rm h}03^{\rm m}15^{\rm s}\!\!.617$&(3)\\
Decl. (ICRS)                                     &$+08\degr52\arcmin26\farcs00$          &(3)\\
Position angle ($^{\circ}$)                      & 128 (or 308$^\star$)                &(4)\\
Inclination angle ($^{\circ}$)                   &  45                &(4)\\
%Systemic velocity (\kms)                         &  4925            &(4)\\
Angular-size distance (Mpc)                   &  68.4             &(3, 5, 6, 7)\\
Luminosity distance (Mpc)                      &  70.8             &(7)\\
Comoving radial distance (Mpc)             &  69.6             &(7)\\
Redshift					             &  $0.0163$     &(7)\\
Linear scale (pc arcsec$^{-1}$)              & 330               &(7)\\
$\log L_{\rm 2-10\;keV}$ (erg s$^{-1}$)    & 43.17          &(8)\\
$\log L_{\rm 14-195\;keV}$ (erg s$^{-1}$)& 43.60          &(8)\\
$\log L_{\rm 8.4\;GHz}$ (erg s$^{-1}$)     & 36.96          &(9)\\
$\log L_{\rm 8-1000\;\mu m}$ (erg s$^{-1}$)& 44.58      &(10)\\
$\log L_{\rm bol.}$ (erg s$^{-1}$)               & 45.30         &(11)\\
Bulge stellar mass (\Msun)                &($1.1\pm0.3)\times10^{11}$&(12)\\
Total gas mass (\Msun)                     &$1.0\times10^{10}$  &(6)\\
$M_{\rm CND,\ total}$ (\Msun)          &$2.7\times10^9$      &(4, 5)\\
$M_B$ (mag)                                    &$-20.9\pm0.2$      &(12)\\
$M_R$ (mag)                                    &$-22.08\pm0.75$      &(13, 14)\\
Velocity dispersion (\kms)                 &$152\pm16$             &(15)\\
$\langle SFR \rangle$ (\Msun~yr$^{-1}$ kpc$^{-2}$)& 50--100&(4)\\
Stellar age (Myr)                               & 110--190                  &(4)\\
RM-based $M_{\rm BH}$ (\Msun)    &$(1.0\pm0.1)\times10^7$&(16, 17, 18)\\
$[{\rm CI}]$(1-0)-based $M_{\rm BH}$ (\Msun) &$1.8^{+2.7}_{-1.1}\times10^7$&(19)\\
\hline
\end{tabular}
\parbox[t]{0.472\textwidth}{\textit{Notes:} (1): \citet{Knapen00}; (2): \citet{Osterbrock93}; (3): I20; (4): D04; (5): \citet{Izumi15}; (6): \citet{Meixner90};  (7) \url{http://www.astro.ucla.edu/~wright/CosmoCalc.html}; (8): \citet{Liu14}; (9): \citet{Perez-Torres09}; (10): \citet{Sanders03}; (11): \citet{Kaspi00}; (12): \citet{Wandel02}; (13): \citet{McLure00}; (14): \citet{McLure01}; (15): \citet{Onken04}; (16): \citet{Peterson14}; (17): \citet{Wang14}; (18): \citet{Peterson04}; (19): this paper. $^\star$: The difference in 180$^\circ$ of the PA will flip the CND's velocity-position diagram but does not change the dynamical results. }
\label{ngc7469property}
\end{table}
%%%%%%%%%%%%%%%%%%%%%%%%%%%%%%%%%%%%%%%%%%%%%%

%%%%%%%%%%%%%% TABLE 2 HST observations %%%%%%%%%%%%%%%%%
\begin{table*}
\caption{\hst/WFC3 UVIS-FIX and ACS WFC images}
\begin{tabular}{cccccccccc}
\hline\hline   
Filter& Camera & Aperture &      UT Date   &   PID  &  PI &  Pixel-scale  &  Exposure time &Zero-point$^a$&$A_\lambda^b$\\
      &        &          &                &        &     &($\arcsec$/pix)&     (s)    &     (Vega mag)    &    (mag)    \\
 (1)  &  (2)   &   (3)    &       (4)      &   (5)  & (6) &      (7)      &     (8)    &      (9)     &     (10)    \\
\hline
F547M &  WFC3  &UVIS2-FIX &2009 November 11&GO-11661&Bentz&     0.04      &$3\times370$&    24.748    &     0.075   \\
F814W (saturation)&  ACS&   WFC    &2006 June 12    &GO-10592&Evans&     0.05      &$2\times200$&    25.943    &     0.105   \\
\hline
\end{tabular}
\label{tab_hst}
\end{table*}
%%%%%%%%%%%%%%%%%%%%%%%%%%%%%%%%%%%%%%%%%%%%%%

%%%%%%%%%%%%%%% TABLE 3 ALMA observations %%%%%%%%%%%%%%%
\begin{table*}
\caption{ALMA observations of the molecular-$^{12}$CO(1-0) and atomic-[CI](1-0) emission lines}
\begin{tabular}{ccccccccccc}
\hline\hline   
  Band   &Emission&   Frequency  &   BMAJ    &    BMIN   &   BPA    &   RMS  &$V_{\rm sys}$&Velocity width& LAS      & Velocity res.\\
              &         &       (GHz)     &($\arcsec$/pc)&($\arcsec$/pc)&($^\circ$)&(mJy beam$^{-1}$ \kms)&   (\kms)    & (\kms)  &($\arcsec$)&     (\kms)   \\
  (1)       &     (2)      &       (3)          &     (4)   &     (5)   &   (6)    &   (7)  &   (8)   &      (9)    &   (10)    &   (11)          \\
\hline 
3           &$^{12}$CO(1-0)    &    115.271   &0.37/123& 0.27/90  & $-$72.5  &34&  4850   &     400     &   3.05  &5.2\\
%cont.     & 2.6 mm &(112.5--116.3)&    --    &      --   &   --     & 47$^b$ &   --    &      --     &   --      & 1.9      &5.1\\
%\hline
8           &  [CI](1-0)  &    492.161   &0.34/113 & 0.31/103 &   75.7   &227&  4831   &     390     &   4.53   &2.5\\
%cont.     & 0.6 mm &(469.3--483.3)&   --      &     --    &   --     & 70$^b$ &   --    &      --     &   --      & 3.9      &2.5\\
\hline
\end{tabular}
\parbox[t]{0.92\textwidth}{\textit{Notes:}{ (1)--(3): ALMA observed bands, emission lines, and their rest frequencies. (4)--(6): The synthesised beam sizes (both in arcsec and parsec) and orientations. BMAJ, BMIN and BPA are the major axis, minor axis and position angle of the beam. (7): The root-mean-squares ({\tt rms}) noise. (8)--(10): The systemic velocity, total velocity width of the whole CND and  the largest angular size of each emission line. (11): The velocity resolution of the emissions.}}
\label{tab_alma}
\end{table*}
%%%%%%%%%%%%%%%%%%%%%%%%%%%%%%%%%%%%%%%%%%%%%%

%%%%%%%%%%%%%% HST Image Figure 1 %%%%%%%%%%%%%%%%%%%
\begin{figure*}
    \centering\includegraphics[scale=0.57]{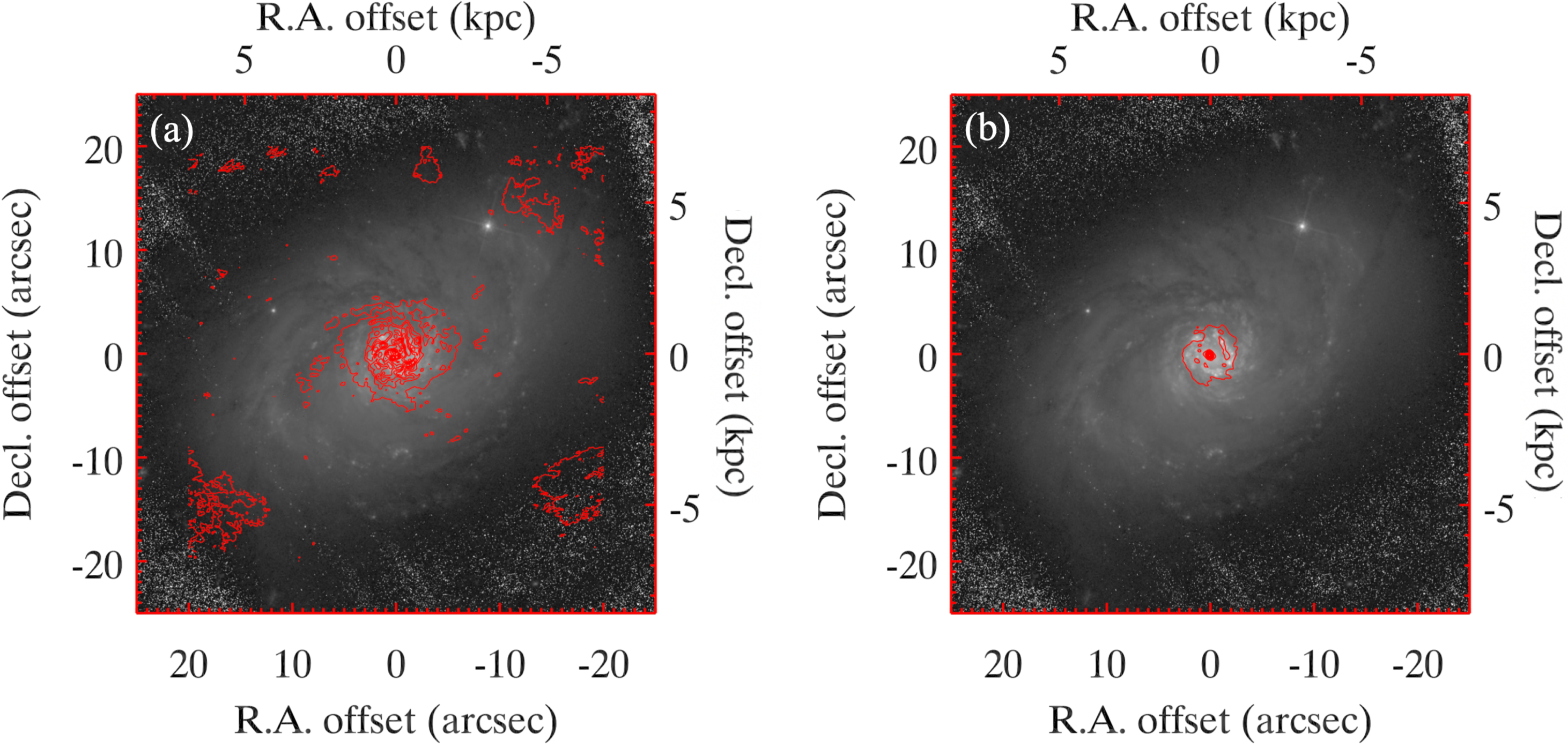}\vspace{5mm}
    \hspace{5mm}
    		  \includegraphics[scale=0.57]{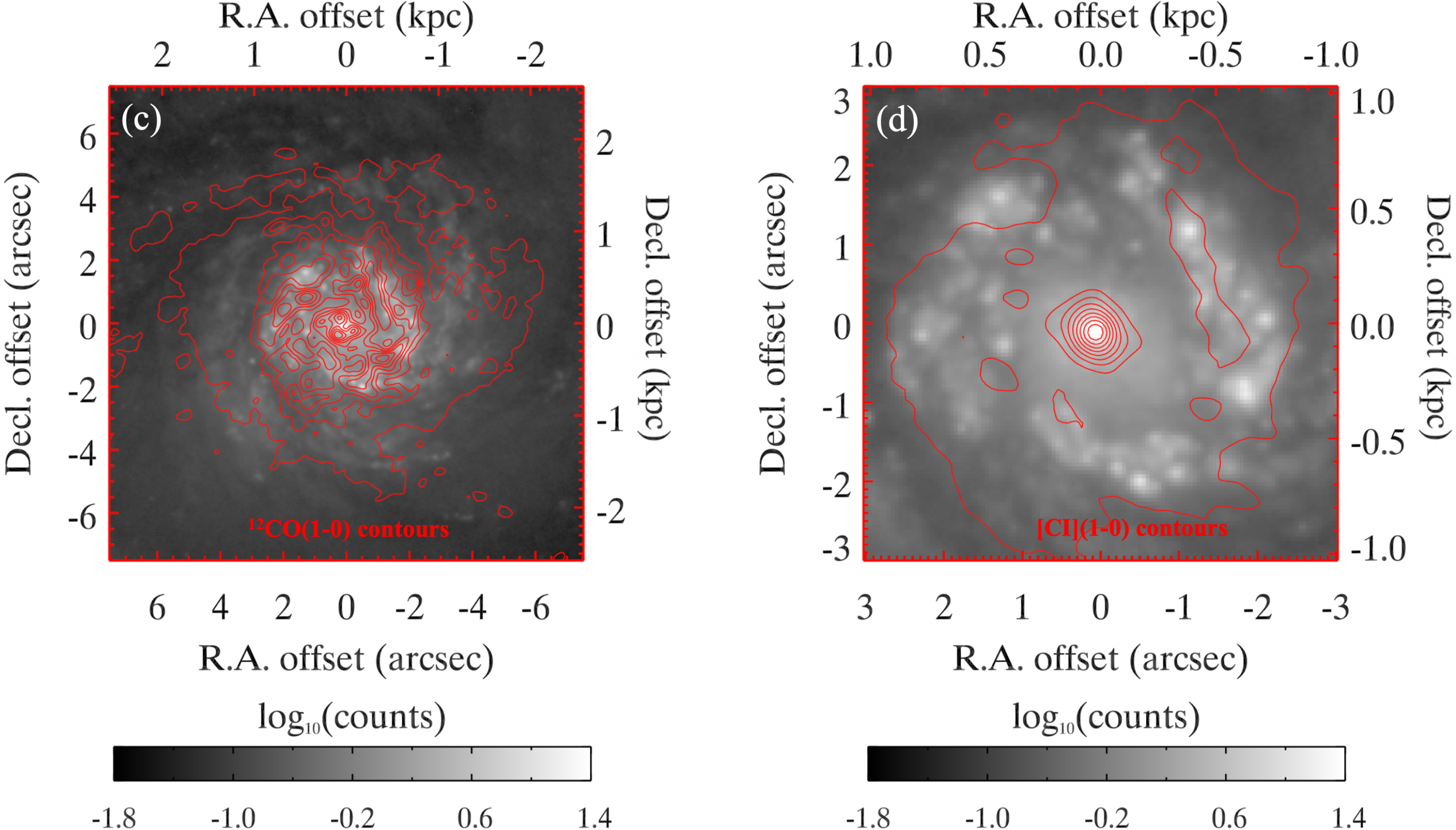}
    \caption{ The \hst/WFC3 F547M image of NGC~7649 within the FOV of $50\arcsec\times50\arcsec$ ($\approx16.5\times16.5$ kpc$^2$) overlaid with the contours of the integrated intensities of molecular-$^{12}$CO(1-0) (panel a) and atomic-[CI](1-0) (panel b) emissions, respectively. These maps show the large scale image and the co-spatial distributions of the dust lanes seen in optical and the molecular/atomic gas seen in mm/sub-mm wavelengths. Here, emission contours are spaced at $0.1\times n\times I_{0,\rm max}$ of $^{12}$CO(1-0) and [CI](1-0) integrated intensities shown in the colour bars of the panels a of Figs~\ref{momsmap1} and \ref{momsmap2}, $n=\overline{1,9}$. In panels c and d, we zoom to the FOV of $15\arcsec\times15\arcsec$ ($\approx5\times5$ kpc$^2$) for $^{12}$CO(1-0) and $6\arcsec\times6\arcsec$ ($\approx2\times2$ kpc$^2$) for [CI](1-0) to show the detailed distributions of the gas emissions at the centre of NGC~7469}. 
    \label{hstalma}   
\end{figure*}
%%%%%%%%%%%%%%%%%%%%%%%%%%%%%%%%%%%%%%%%%%%%

%%%%%%%%%%%%%%%%%%%%%%%%%%%%%%%%%%%%%%%%%%%%%%
\section{Data}\label{data}

%%%%%%%%%%%%%%%%%%%%%%%%%%%%%%%%%%%%%%%%%%%%%%
\subsection{\emph{Hubble Space Telescope} (\emph{HST}) images}\label{hst} 

We used the optical \hst\ observations in the WFC3/UVIS-FIX F547M and ACS/WFC F814W filters to create a mass-follows-light model (Section~\ref{massmodels}). The F814W image suffers from central saturation due to the bright AGN, which will be masked out in the subsequent analysis. Also, there is a 20\% difference in the pixel sizes between the two instruments, WFC3 vs. ACS. We thus downloaded the raw {\tt flt} frames of the ACS F814W image from \hst/Mikulski Archive for Space Telescopes (MAST) and re-reduced these images using the \texttt{drizzlepac/Astrodrizzle} package\footnote{\url{https://www.stsci.edu/scientific-community/software/drizzlepac}} \citep{Avila12} to a final pixel scale of $0\farcs04$. More details of these images are shown in Table \ref{tab_hst}.

We aligned these \hst\ images to the galaxy centre determined from I20. Next, we used the \texttt{Tiny Tim\footnote{\url{http://tinytim.stsci.edu/sourcecode.php.}}} routine \citep{Krist95, Krist11} to create point spread function (PSF) models for individual \hst\ exposure frames of the involved filters. Then, we inserted them into the corresponding {\tt flc} images at the galaxy centre in each frame to simulate the observations. The final PSF model is the combination of these PSF models created by \texttt{Astrodrizzle} \citep{Rusli13, denBrok15, Nguyen17, Nguyen18, Thater19a}, which has a FWHM of $0\farcs08$ ($\approx$ 27 pc) and will be used to decompose the mass model (Section~\ref{massmodels}).

Figure~\ref{hstalma} shows the F547M image. There is a starburst ring with bright and resolved super-star clusters distributed around the bright nucleus within the annulus of 1$\farcs$5--2$\farcs$5 ($\approx$ 500--833 pc), as well as prominent dust lanes on the north and northeast sides, extending to a radius of at least $10\arcsec$ ($\approx3.3$ kpc) from the centre.

%%%%%%%%%%%%%%%%%%%%%%%%%%%%%%%%%%%%%%%%%%%%%
\subsection{ALMA observations}\label{almaobs}

 \citet{Izumi20} reported a variety of bright gas emissions in the nucleus of NGC 7469, including molecules (e.g. $^{12}$CO(1-0), $^{12}$CO(2-1), $^{12}$CO(2-1), $^{12}$CO(3-2)) and atomic [CI](1-0). An accurate weighing of the central SMBH requires that these emissions are observed at high spatial resolution, and show bright, compact distributions towards the galaxy centre and regular kinematics in the entire CND. Owing to the XDR effect (i.e. dissociation of CO into C), [CI](1-0) surely probes the close vicinity of this AGN as evidenced by the position-velocity diagram (PVD). We thus can extend this analysis towards higher redshift thanks to the higher rest frequency of [CI], which is much easier than the CO lines observed here. On the other hand, an extended CND traces the total enclosed mass out to the radius covered by the gas, which helps to separate the \Mbh\ from the masses of the remaining galaxy components (i.e. stars + ISM + non-luminous- stellar remnants) better. Regarding this aspect, the $\approx5$\% higher in spatial resolution and more extended of $^{12}$CO(1-0) make it a better transition for dynamical modelling than other CO lines. We thus chose the atomic-[CI](1-0) and molecular-$^{12}$CO(1-0) emissions among these as the best bets to perform \Mbh\ measurement that compromises both these requirements. 

The nuclear molecular-$^{12}$CO(1-0) and atomic-[CI](1-0) emissions of NGC~7469 were observed in bands 3 and 8 of ALMA cycle 5 (PID: 2017.1.00078.S, PI: T. Izumi) at the spatial-resolution scales of $\approx0\farcs3$ (or $\approx100$ pc) that are $\approx55$ times larger than the SOI of the black holes predicted from RM \citep[$\approx0\farcs006$ or $\approx1.9$ pc;][]{Peterson04, Peterson14, Wang14}. These bright and high-quality observations reveal that their kinematics are suitable for gas-dynamical modelling to measure the central \Mbh. Details of data reduction and imaging of these visibilities are presented in I20. In Table~\ref{tab_alma}, we show the observational properties of these two emissions.

We created a three-dimensional (3D) cubes of (R.A., decl., velocity) for these $^{12}$CO(1-0) and [CI](1-0) emissions using the Common Astronomy Software Applications \citep[{\tt CASA};][package version 5.4]{McMullin07}. The $^{12}$CO(1-0) and [CI](1-0) integrated intensities are shown in contours in Fig.~\ref{hstalma}, overlaid on the F547M image. It is clear that the atomic-[CI](1-0) emission is more compact and strongly peaked toward the centre ($<$ 5$\arcsec$ or 1.7 kpc) than the molecular-$^{12}$CO(1-0) emission when we solely look at the central kpc regions (i.e.\ inside the starburst ring). However, the molecular-$^{12}$CO(1-0) emission shows a more extended distribution ($\approx7\arcsec$ or 2.3 kpc) with some fainter features distributed further out at $\approx$ 15$\arcsec$--20$\arcsec$ (or $\approx$ 5--6.6 kpc).   

%%%%%%%%%%%%%%%%% moms maps Figure 3 %%%%%%%%%%%%%%%%%
\begin{figure*}
    \centering\includegraphics[scale=0.8]{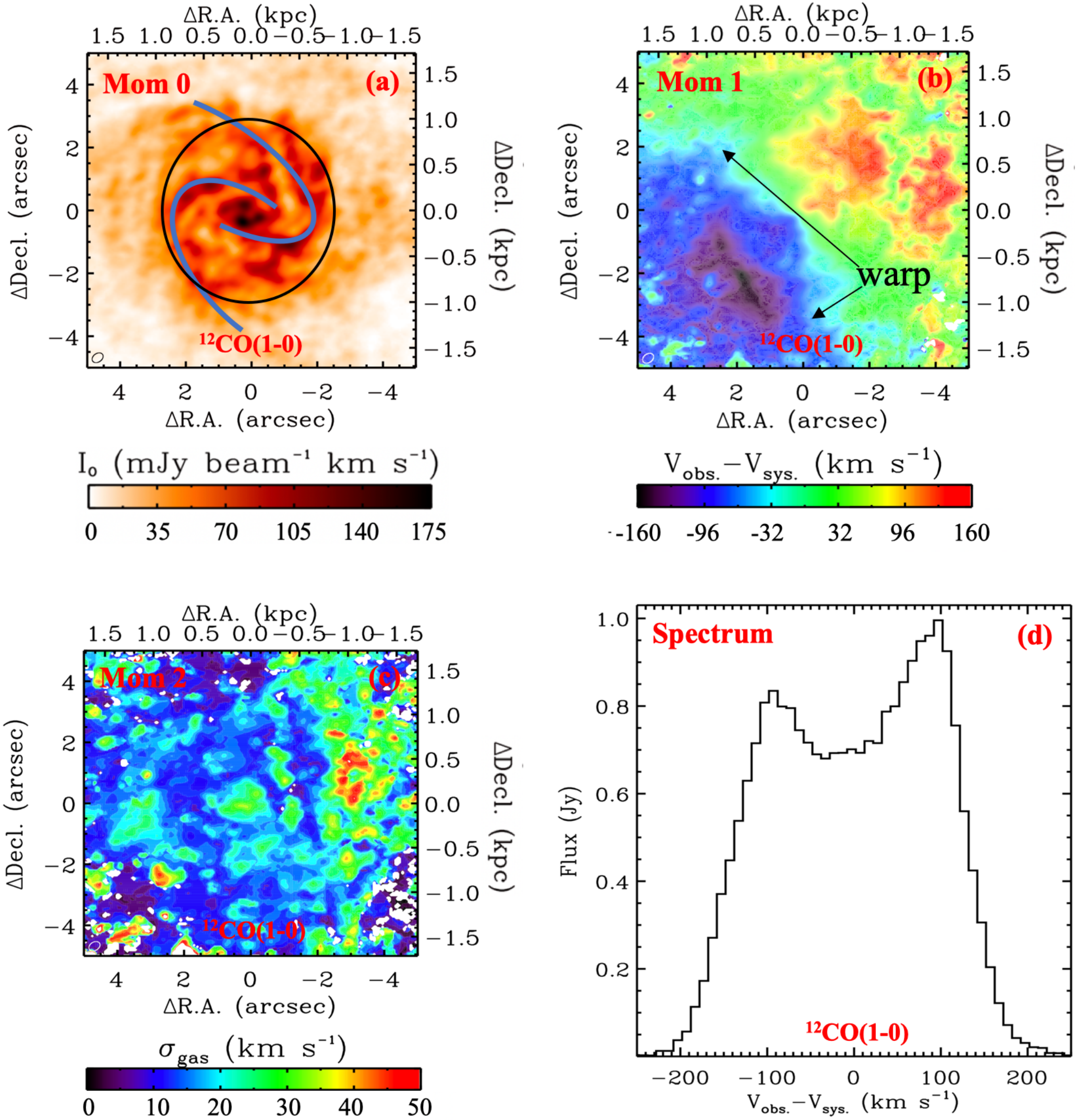}
    \caption{The moment maps of $^{12}$CO(1-0) show in the order of the integrated intensity (panel a), the intensity-weighted mean LOS velocity field (panel b), and the intensity-weighted LOS velocity dispersion (panel c) within the FOV of $10\arcsec\times10\arcsec$ ($\approx3.3\times3.3$ kpc$^2$). In the moment 0 map, the black circle shows the position of the starburst ring \citep{Soifer03, Diaz-Santos07}, while the cyan arcs highlight the two-arm spiral structure of the CND. In the moment 1 map, there is a small warp elongated along the minor axis, creating a twist seen in this direction. In the moment  2 map, the high-intensity-weighted LOS velocity dispersion regions are co-spatial with the high-surface-brightness density regions in the moment 0 map within the region of $2\arcsec$ ($\approx660$ pc). Further west of this region where the gas distributes regularly in the CND, the LOS velocity dispersion is also high and peaks at the region where the gas is faint (mom 0) and the LOS velocity field drops suddenly in the redshifted side (mom 1) due to low signal-to-noise. The synthesised beam size listed in Table~\ref{tab_alma} is shown as an ellipse at the bottom left of each panel. The ($\Delta{\rm R.A.}, \Delta{\rm decl.})=(0,0)$ position on these maps indicates the kinematic/galaxy centre. In panel d, we show the integrated spectrum extracted within the same FOV, showing an asymmetric double-horn shape of a rotating disc.}
\label{momsmap1}   
\end{figure*} 
%%%%%%%%%%%%%%%%%%%%%%%%%%%%%%%%%%%%%%%%%%%%%

%%%%%%%%%%%%%% moms maps Figure 4  %%%%%%%%%%%%%%%%%%%
\begin{figure*}
    \centering\includegraphics[scale=0.8]{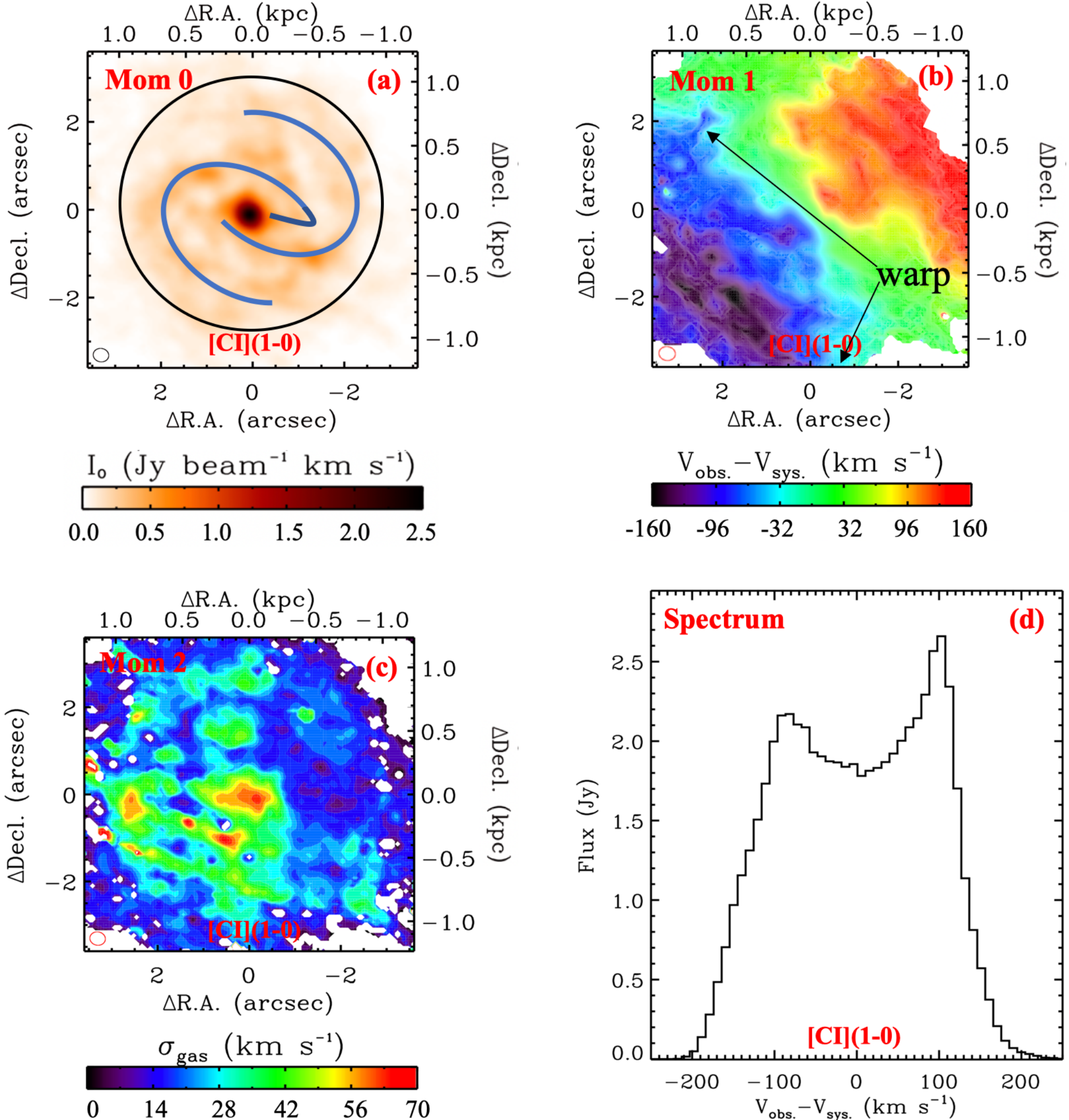}
    \caption{Same as Fig.~\ref{momsmap1} but for [CI](1-0) in a smaller FOV of $7.2\arcsec\times7.2\arcsec$ (or $2.4\times2.4$ kpc$^2$), but some differences are described here.  In the integrated intensity (panel a) map, the black circle and blue spiral arms illustrate the same region and the arcs of the two-arm spiral structure of the CND as similar as those in the panel a of Fig.~\ref{momsmap1}, although we show them in a smaller FOV of the map (i.e. their illustrations are bigger in sizes). In the moment 2 map (panel c), there is an opposite distribution of the atomic-[CI](1-0) gas compared to the molecular-$^{12}$CO(1-0) gas in Fig.~\ref{momsmap1}.  The high-velocity-dispersion regions are mostly located in the south and southeast sides of the nucleus not in the north and northwest.  The reason for these inverse velocity-dispersion distributions is currently unknown. Future works of both atomic and molecular gas surveys in a sample of AGN will disentangle this issue better.} 
\label{momsmap2}   
\end{figure*}
%%%%%%%%%%%%%%%%%%%%%%%%%%%%%%%%%%%%%%%%%%%%%

We also created the integrated intensity (moment 0), intensity-weighted mean LOS velocity field (moment 1), and intensity-weighted LOS velocity dispersion (moment 2) maps for $^{12}$CO(1-0) and [CI](1-0) using the moments masking technique \citep{Dame01, Dame11} and showed them in the panels a, b, and c of Figs \ref{momsmap1} and \ref{momsmap2}, respectively. To do this, we spatially smoothed each channel by a factor of $\alpha$ then varied $\alpha$ and gauged the spatial and velocity coherence of the signal. The spatial smoothing increases the sensitivity but decreases the angular resolution, helping to suppress noise peaks. Next, we performed  ``$\beta\sigma$-clipping'', where $\beta$ is a positive factor and $\sigma$ is the {\tt RMS} noise (recorded in column 7 of Table~\ref{tab_alma}). Thus, we decided all channels with intensities  $<\beta\sigma$ were set to zero at any given spatial position. This mask created from the smoothed cube was only used to identify and mask out emission-free regions of the original cube, while the moment maps with full spatial and velocity resolutions were created from the original cube still. We varied $\alpha$ and $\beta$ for their appropriate choices to obtain the best moment maps, which yields $\alpha=3$ and $\beta=0.5$. 

%%%%%%%%%%%%%% moms maps Figure 4  %%%%%%%%%%%%%%%%%%%
\begin{figure}
    \centering\includegraphics[scale=0.33]{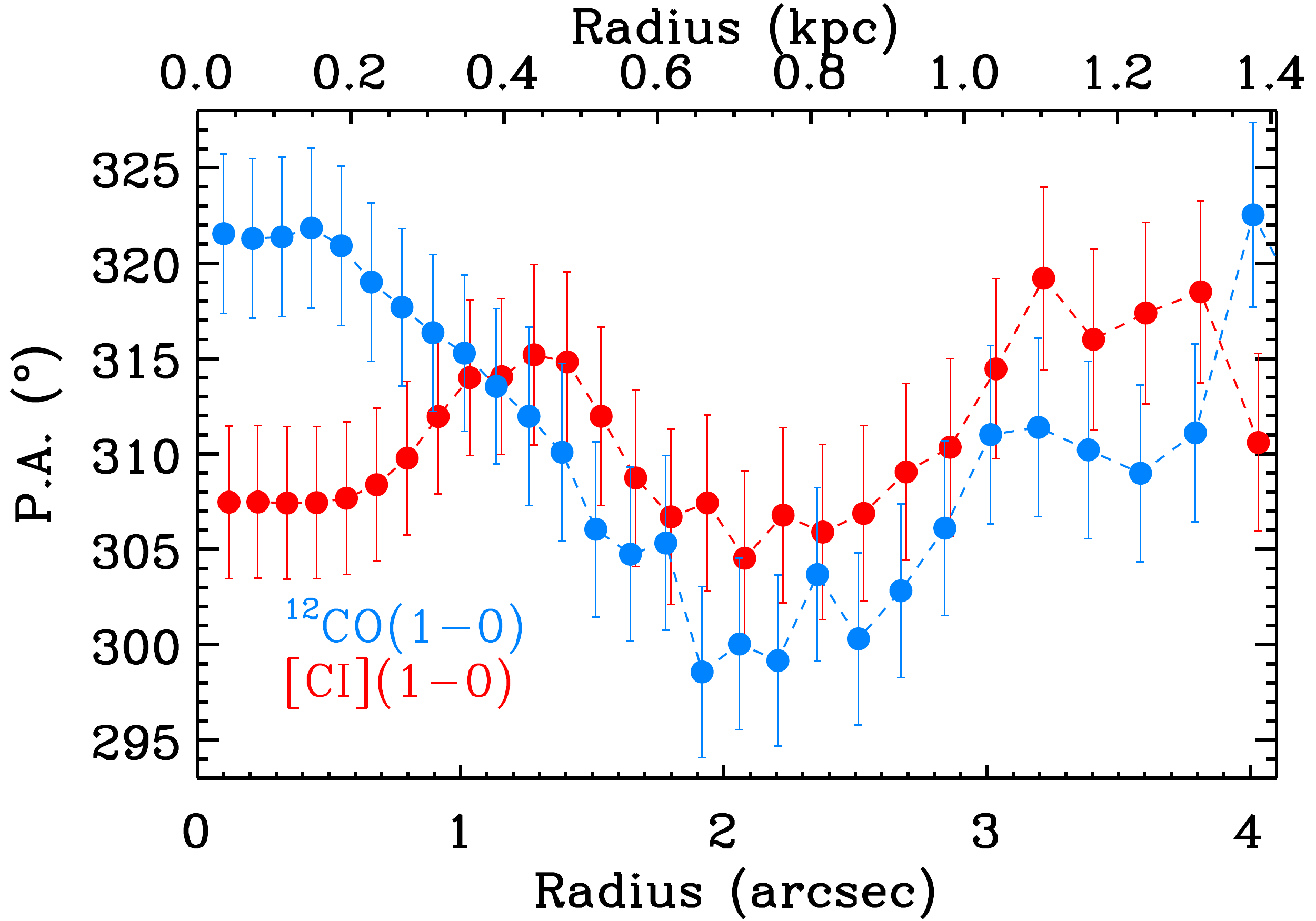}
     \caption{The radial position angle (PA) profiles of the $^{12}$CO(1-0) (cyan) and  [CI](1-0) (red) emission lines extracted along the galaxy major axis using the {\tt Kinemetry} code \citep{Krajnovic06}}.
    \label{radialPA} 
\end{figure}
%%%%%%%%%%%%%%%%%%%%%%%%%%%%%%%%%%%%%%%%%%%%%

The gas have ring-like structures with diameters of $\approx$ 7$\arcsec$ for $^{12}$CO(1-0) and $<$ 5$ \arcsec$ for [CI](1-0), indicating by features of two-arm/bi-symmetric spiral structures (showing as the two cyan arcs in the panels a of Figs~\ref{momsmap1} and \ref{momsmap2}). In the molecular-$^{12}$CO(1-0) gas integrated intensity map (panel a of Fig.~\ref{momsmap1}), the emission is resolved with the three high-surface-brightness regions within $0\farcs5$ and other high-surface-brightness areas associated with the star-forming ring (black circle).  Whereas, the atomic-[CI](1-0) gas emission is centrally peaked within central $0\farcs5$, and the two-arm spiral structure of the high-surface-density regions opens a bit narrower than that of $^{12}$CO(1-0) (panel a of Fig.~\ref{momsmap2}).

The intensity-weighted mean LOS velocity fields, which are beam-convolved, of both emission lines reveal a rotating disc with a total velocity width ($\Delta V$) of $\approx400$ \kms\ for $^{12}$CO(1-0) and $\approx390$ \kms\ for [CI](1-0) and small warps seen along the minor axis (panels b of Figs~\ref{momsmap1} and \ref{momsmap2}). These warps would make our dynamical modellings become complicated and hard to measure the \Mbh. Panels c of Figs~\ref{momsmap1} and \ref{momsmap2} show the intensity-weighted mean LOS velocity dispersion ($\sigma_{\rm gas}$) maps of these emissions. The $\sigma_{\rm gas}$ shows spaxel variations within the ranges of $\approx(3$--$51)$ \kms\ for $^{12}$CO(1-0)  and $\approx(3$--$67)$ \kms\ for [CI](1-0). Generally, the high intensity-weighted mean LOS velocity dispersion regions are co-spatial with the high-surface-brightness density regions, i.e. the centrally peaked region, star-forming ring, and regions associated with the two-arm spiral structure. It is notable that at large radii ($r\gtrsim2\arcsec$ or 660 pc), these high-velocity-dispersion regions are highly dominant in the west and northwest sides of the $^{12}$CO(1-0) map, while in the [CI](1-0) map they are mostly located in the south and southeast sides. The origin of these inverse distributions of the intensity-weighted mean LOS velocity dispersions between these two emission lines is unknown. However, they do not affect the dynamical modelling of \Mbh.  

In panels d of Figs~\ref{momsmap1} and \ref{momsmap2}, we show the integrated spectrum of the $^{12}$CO(1-0) and [CI](1-0) CNDs created by extracting the fluxes within the nuclear regions of $10\arcsec\times10\arcsec$ ($\approx3.3\times3.3$ kpc$^2$) and $7\farcs2\times7\farcs2$ ($\approx2.4\times2.4$ kpc$^2$) of the corresponding datacubes. These profiles show a classical double-horn shape of a rotating disc but asymmetric with some missing blue-shifted velocity channels. 

Similarly, Fig. \ref{almapvd} shows the PVDs of $^{12}$CO(1-0) and [CI](1-0) extracted from a cut of three pixels in width ($0\farcs81\approx270$ pc) through the major axis (${\rm PA}=308^{\circ}$). There are sharp increases of the rotations towards the galaxy centre in both emissions, interpreted as the Keplerian motions (i.e. $V_{\rm circ}(r)\propto r^{-0.5}$) caused by either a SMBH or a centrally massive and compact cluster of baryonic matter residing at the heart of NGC~7469. The asymmetry in the CND is only seen in the blue-shifted Keplerian motion of the $^{12}$CO(1-0) PVD with $\approx20$ \kms\ less than the red-shifted Keplerian motion. This asymmetric emission distribution in the PVD may be caused by a smaller fraction of molecular gas mass on the blueshifted component than that of the redshifted component. Other possibility is that they have different excitation  condition \citep{Imanishi18, Izumi18}. Here, we rule out the possibility of dust extinction because there is less extinction in this blueshifted component (Section~\ref{colourmass}). However, this rotating-asymmetric feature is not present in the [CI](1-0) PVD, suggesting this emission line is actually a more appropriate line to measure the \Mbh\ than the molecular-$^{12}$CO(1-0) line for this AGN NGC~7469. This argument is strengthened when we examine these emissions in the galaxy-minor-axis (${\rm PA}=218^{\circ}$) PVDs as shown in Fig.~\ref{pvdminor}. The [CI](1-0) emission is symmetric, while the $^{12}$CO(1-0) emission is asymmetric with missing the blueshifted motions towards the galaxy centre. On the other hand, the latter is more extended towards larger radii than the former. These features again tell us that [CI](1-0) is a best bet to estimate the centrally compact mass (i.e.\ black hole mass), while the $^{12}$CO(1-0) line is preferable in tracing the enclosed mass (i.e.\ mass-to-light ratio, \ml) of NGC~7469 out to $\approx7\arcsec$ \citep[see panel a of Fig.~\ref{hstalma}; also see the usage of $^{12}$CO(2-1) in][]{Izumi20}. Importantly, at the spatial scales measured in this work, there are somewhat insignificant non-circular motions (i.e. outflows or inflows $\approx20\pm10$ (kinematic uncertainty) \kms; will be discussed later in Section~\ref{defaultresult}) are found in both lines.

%%%%%%%%%%%%%%%%%%%%%%%%%%%%%%%%%%%%%%%%%%%%%%%%%%%%%%%
%%%%%%%%%%%%%%%%%%%%%%%%%%%%%%%%%%%%%%%%%%%%%%%%%%%%%%%
\section{Dynamical Models}\label{kinms}

In this section, we present the KinMS tool \citep[Section~\ref{modeldescription};][]{Davis14} and the {\it default-mass model} (Section~\ref{massmodels}) of NGC~7469 used to estimate the central \Mbh. Next, we report the results in Section~\ref{defaultresult}. Alternatively, we test our dynamical modellings with various galaxy mass models based on different assumptions of colour variability and colour--\ml\ correlations \citep[][hereafter B01; N19]{Bell01, Nguyen19a} in Section \ref{colourmass}. A simpler version of this analysis is presented in I20 as well to dynamical measure the CO-to- or [CI]-to-H$_2$ mass conversion factors.

%%%%%%%%%%%%%%%%%%%%%%%%%%%%%%%%%%%%%%%%%%%%%%%%%%
\subsection{KinMS model}\label{modeldescription}

The KinMS tool optimises a datacube in two steps. First, it creates a simulated cube with a given set of model parameters for comparison to observables. Second, it determines the best-fitting model by using an efficient method to explore the parameter-space. 

To create a simulated cube, KinMS adopts a parametric function describing the distribution and kinematics of molecular gas. The gas is assumed to move on circular orbits governed by a circular velocity curve, calculated from the {\tt mge\_circular\_velocity} procedure within the Interactive Data Language (IDL) Jeans Anisotropic Modelling \citep[JAM\footnote{\url{https://purl.org/cappellari/software}};][]{Cappellari08} package. This circular velocity curve is then used as the corresponding input of the axisymmetric-mass model specified via the multi-Gaussian expansion \citep[MGE;][]{Emsellem94a, Cappellari02} parametrisation, including all mass components: stars, the interstellar medium (ISM; gas + dust) and a putative SMBH \citep{Davis13}.

The KinMS tool then uses this simulated cube for optimising the data cube. Specifically, it utilises the {\tt emcee} technique \citep{Foreman-Mackey13},  the affine-invariant ensemble sampler based on the Markov Chain Monte Carlo (MCMC) method, and the Bayesian analysis framework to walk through the parameter space \citep{Goodman10}. The model calculates $\chi^2$ at every step and uses it to determine the next move until reaching the minimum-$\chi^2$. In practice, this is all accomplished by using the {\tt Python} code {\tt KINMSpy\_MCMC}\footnote{\url{https://github.com/TimothyADavis/KinMSpy_MCMC}}. To find the best-fit parameters and errors, we determine the probability distribution function (PDF) likelihood calculated from the full calculation pool for the posterior distribution. 

Dynamical pressure distributed by the highly turbulent $\sigma_{\rm gas}$ of a thick disc could potentially cause bias in the \Mbh\ because the gas pressure supports the gas against gravity \citep{Binney87}. Observations with a high ratio of $\sigma_{\rm gas}/V_{\rm rot}$ will spoil the thin disc assumption of gas moving on purely circular orbits \citep{Coccato06, Barth16a, Boizelle19}. Using the dynamics of [CI](1-0) and $^{12}$CO(2-1), I20 found $\sigma_{\rm gas}/V_{\rm rot}\approx0.35$ ($r\lesssim0\farcs5$) and $\sigma_{\rm gas}/V_{\rm rot}\approx0.19$ ($r>0\farcs5$) (see panels a and b of their Fig.~11), which indicates a dynamically thin ($d_{\rm t}\approx0$) and cold ($V_{\rm rot}\approx V_{\rm circ}$) CND for NGC~7469. That means the dynamically asymmetric drift from turbulent velocity dispersion of the gas can be ignored \citep[][Nguyen et al. submitted]{Davis20}.

The KinMS tool also uses a radial parametric function to describe gas surface brightness distribution (SB). Here, we assumed the axisymmetric SBs for [CI](1-0) and $^{12}$CO(1-0) and modelled them using a Gaussian. We identified the Gaussian centre to be the kinematic centre ($x_{\rm c}$, $y_{\rm c}$) and incorporated their normalization factors into the total flux parameters ($F$). Thus, we described the $^{12}$CO(1-0) and [CI](1-0) SBs with only one free parameter: the dispersion of the Gaussian ($G_\sigma$).

Additionally, to account for the small warps seen in both intensity-weighted mean LOS velocity fields of $^{12}$CO(1-0) and [CI](1-0) (panels c of Figs~\ref{momsmap1} and \ref{momsmap2}), we extracted their radial PA profiles along the major axis via the \texttt{Kinemetry}\footnote{\url{http://davor.krajnovic.org/idl/\#kinemetry}} code \citep{Krajnovic06}, then used them as an additional input in the same manner of using the circular velocity curve. These PA profiles vary in the range of (295--325)$^\circ$ for $^{12}$CO(1-0) and (300--320)$^\circ$ for [CI](1-0) across $4\arcsec$ as shown in Fig.~\ref{radialPA}. Accounting for PA variations in our dynamical models help to mimic the warps seen along the galaxy-minor axis of the velocity-field maps of $^{12}$CO(1-0) (panel b of Fig.~\ref{momsmap1}) and [CI](1-0) (panel b of Fig.~\ref{momsmap2}). Dynamical models with constant PAs will create  axisymmetric kinematic maps that cannot well reproduce the data, which are asymmetric along the minor axis. This mismatch between the data and the model biases the \Mbh\ estimates and produces some large residuals or artificial non-circular motions in the {\tt Data--Model} maps of the velocity fields. 

Overally, the KinMS model matches the observations with nine free parameters, including $G_\sigma$, $i$, $F$, $x_{\rm c}$, $y_{\rm c}$, $M_{\rm BH}$, \ml, $\sigma_{\rm gas}$, and a velocity offset ($v_{\rm off}=V-V_{\rm sys}-V_{\rm rot}$) of the kinematic centre relative to the rotation and the systemic velocity of the whole galaxy.

%%%%%%%%%%%%%%%%%%%%%%%%%%% PVD major Figure 5 %%%%%%%%%%%%%%%%%%
\begin{figure*}
    \centering\includegraphics[scale=0.58]{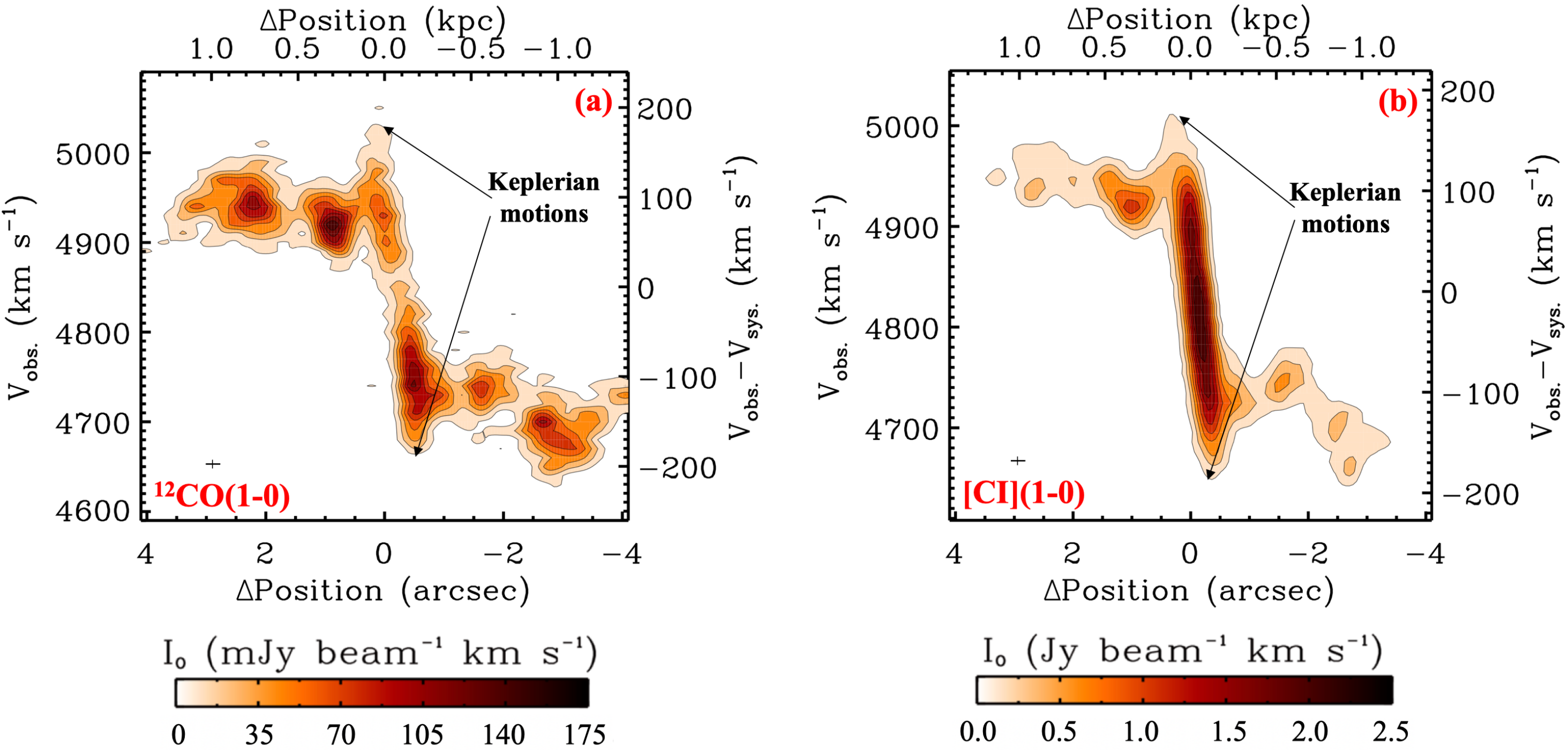}
    \caption{The PVDs extract along the major axis (${\rm PA}=308^{\circ}$) in the FOVs of $10\arcsec\times10\arcsec$ ($\approx3.3\times3.3$ kpc$^2$) for $^{12}$CO(1-0) (panel a) and $7.2\arcsec\times7.2\arcsec$ ($\approx2.4\times2.4$ kpc$^2$) for [CI](1-0) (panel b) with a slit of three pixels in width ($0\farcs81$ or 270 pc). Keplerian motions towards the galaxy centre present in both diagrams. However, that motion of [CI](1-0) is more symmetric than $^{12}$CO(1-0). Small pluses at the bottom left corners show the errors of our measurements. Contour levels are plotted at $n\times0.1\times I_{0,\rm max}$ of $^{12}$CO(1-0) and [CI](1-0) shown in the colour bars (also Figs~\ref{momsmap1} and \ref{momsmap2}), $n=\overline{1,9}$. It is clear that the atomic-[CI](1-0) emission is much more centrally-concentrated than the molecular-$^{12}$CO(1-0) emission. }
\label{almapvd}   
\end{figure*}
%%%%%%%%%%%%%%%%%%%%%%%%%%%%%%%%%%%%%%%%%%%%%%%%%%%%%%%%%

%%%%%%%%%%%%%%%%%%%%%%%%%% PVD minor Figure 6 %%%%%%%%%%%%%%%%%
\begin{figure}
    \centering\includegraphics[scale=0.3]{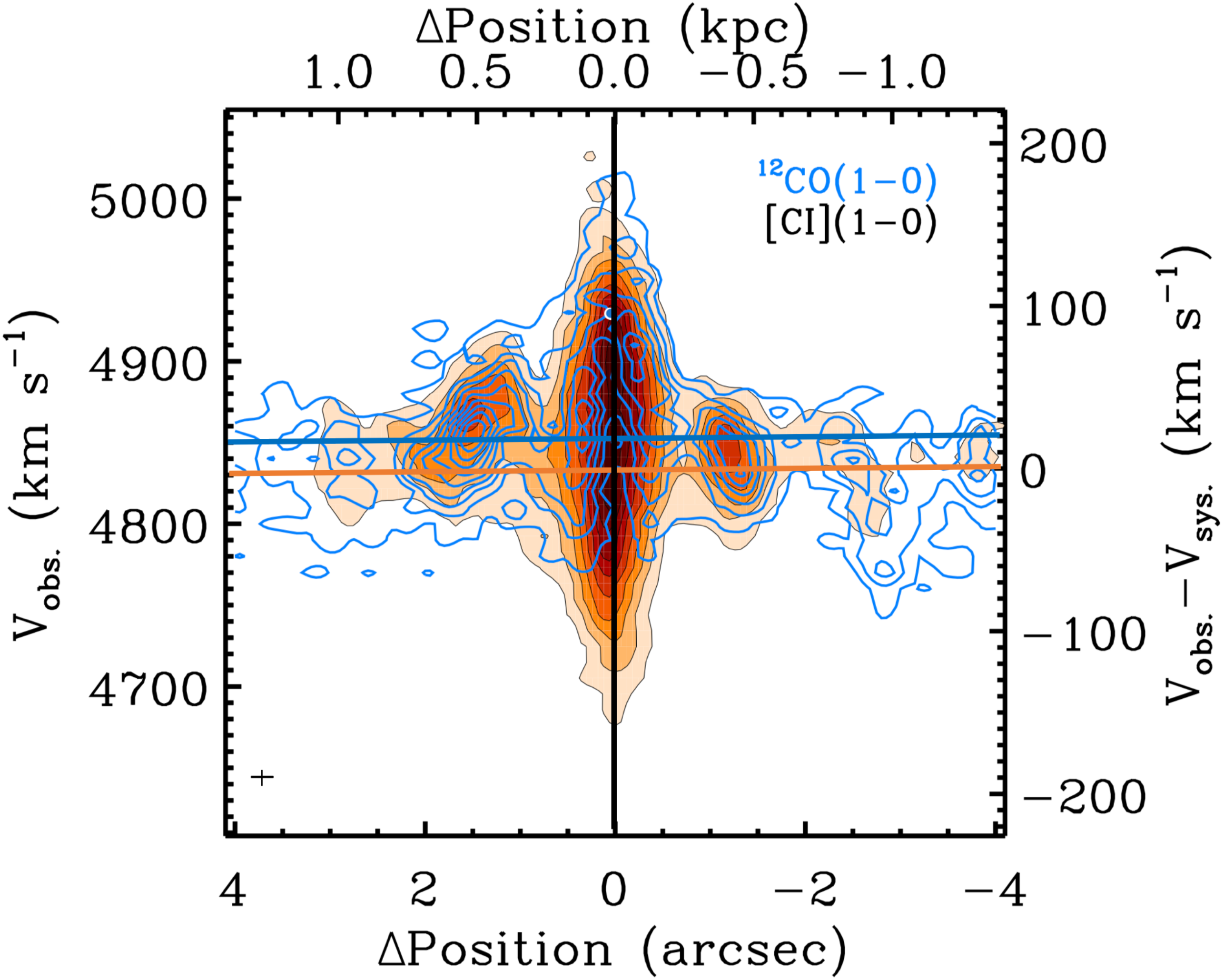}
    \caption{Same as Fig. \ref{almapvd} for the PVDs extract along the minor axis (${\rm PA}=218^{\circ}$). There are no circular motions in both emissions. However, the PVD of [CI](1-0) (red) is more symmetric than the PVD of $^{12}$CO(1-0) (blue). The horizontal lines with corresponding colours indicate the systemic velocities of [CI](1-0) and $^{12}$CO(1-0), while the vertical black line anchors the emission centre.} 
\label{pvdminor}   
\end{figure}
%%%%%%%%%%%%%%%%%%%%%%%%%%%%%%%%%%%%%%%%%%%%%%%%%%%%%%

%%%%%%%%%%%% HST F547M PSF table 5 %%%%%%%%%%%%%%%%%%
\begin{table}
\caption{MGE Parameters of the \hst/WFC3 F547M PSF}
\begin{tabular}{cccc}
\hline\hline   
$j$&Total Count of Gaussian$_j$&$\sigma_j$& $q_j$\\
   &                           & (arcsec) &  \\
(1)&(2)& (3) &(4)\\
\hline
1&0.18&0.02&0.99\\
2&0.64&0.06&1.00\\
3&0.07&0.18&0.99\\
4&0.06&0.37&0.99\\
5&0.02&0.92&0.98\\
6&0.03&1.55&0.98\\
\hline
\end{tabular}
\parbox[t]{0.472\textwidth}{\textit{Notes:}{ Total count, width ($\sigma_j$) and axis ratio ($q_j$) of each Gaussian $j$.}}
\label{tab_psfmge}
\end{table}
%%%%%%%%%%%%%%%%%%%%%%%%%%%%%%%%%%%%%%%%%%%%%%

%%%%%%%%% HST F547M light MGE Table 6 %%%%%%%%%%%% 
\begin{table}
\caption{The \hst/WFC3 F547M Stellar-light-MGE Model}
\begin{tabular}{cccc}
\hline\hline   
$j$ & $\log(\Sigma_{\star,j}/{\rm L}_\odot\,{\rm pc}^2)$ & $\sigma_j$ & $q_j$\\
   &     &(arcsec)& \\
(1)&       (2)  &  (3)        &  (4)\\
\hline 
1  & 5.52  & 0.02   & 1.00\\
2  & 4.79  & 0.04   & 1.00\\
3  & 3.44  & 0.27   & 1.00\\ 
4  & 2.82  & 0.96   & 1.00\\ 
5  & 1.73  & 1.94   & 0.86\\ 
6  & 1.45  & 7.60   & 0.62\\ 
7  & 0.34  & 18.1   & 0.82\\ 
\hline
\end{tabular} 
\parbox[t]{0.472\textwidth}{\textit{Notes:}{ Stellar-light-surface density ($\Sigma_{\star,j}$), width ($\sigma_j$) and axis ratio ($q_j$) of each deconvolved Gaussian component $j$.}}
\label{tab_f547mmge}
\end{table}
%%%%%%%%%%%%%%%%%%%%%%%%%%%%%%%%%%%%%%%%%%%%%%

%%%%%%%% HST F547M MGE model Figure 9 %%%%%%%%
\begin{figure*}
    \includegraphics[scale=0.59]{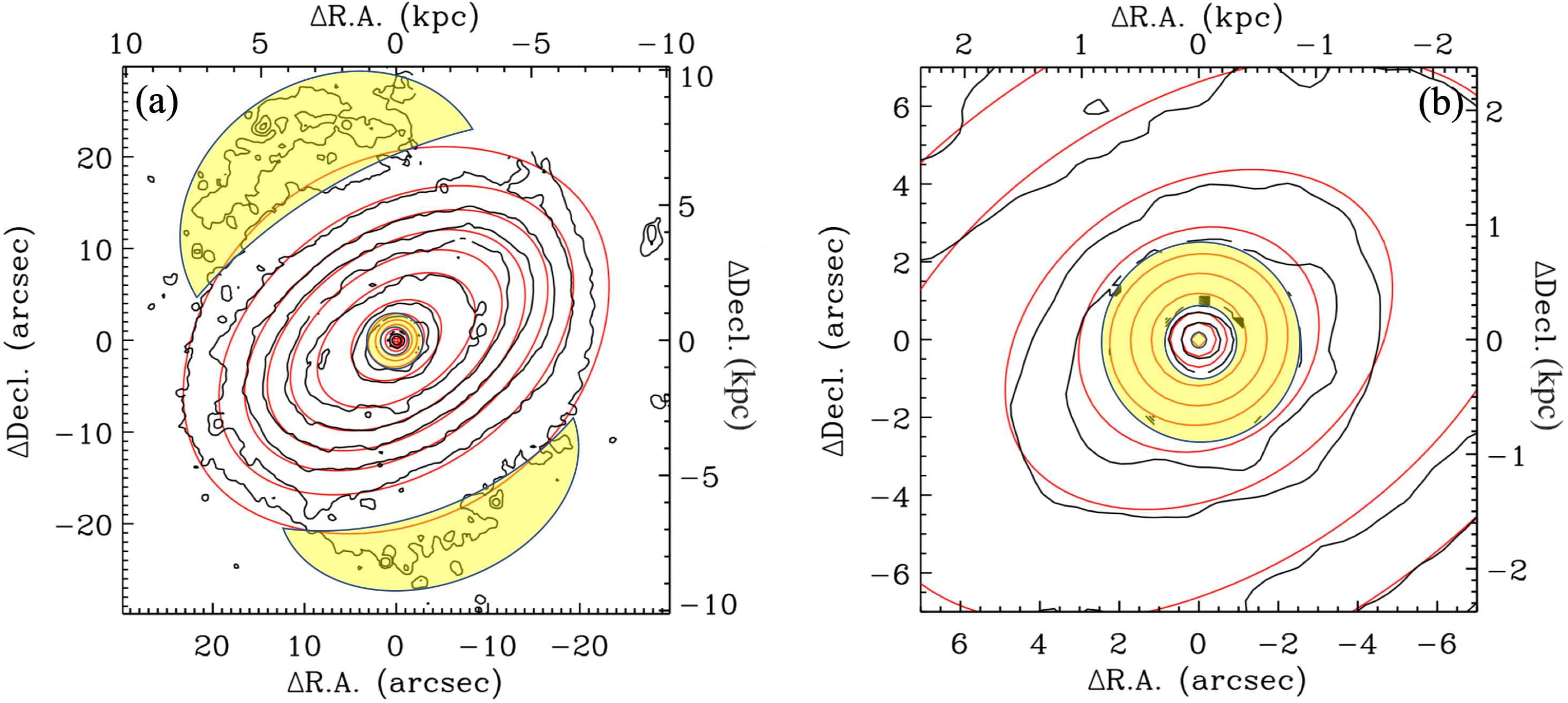}
    \caption{The comparison between the \hst/F547M photometry of NGC~7469 (black) and its corresponding best-fit MGE model (red) at the same radii, which illustrate the whole galactic scale $60\arcsec\times60\arcsec$ ($\approx20\times20$ kpc$^2$, panel a) and the zoom-in of $14\arcsec\times14\arcsec$ ($\approx4.6\times4.6$ kpc$^2$, panel b) FOV. Yellow regions are the pixel excluded during the MGE fit.}  
\label{hstF547mmge}
\end{figure*}
%%%%%%%%%%%%%%%%%%%%%%%%%%%%%%%%%%%%%%%%%%%%%%%

%%%%%%%%%%%%%%%%%%%%%%%%%%%%%%%%%%%%%%%%%%%%%%
\subsection{The default-galaxy-mass model}\label{massmodels}

{\it Stellar mass: } We applied the \texttt{IDL mge\_fit\_sectors} routine, version 4.14\footnote{\url{http:purl.org/cappellari/software}} \citep{Cappellari02} of the MGE model, to parametrise the stellar-mass distribution. We first modelled the F547M PSF as circular MGE model and tabulated it in Table~\ref{tab_psfmge}. Then, we included this PSF MGE in the second axisymmetric-MGE fit for the photometric MGE. We also masked out pixels contaminated by the AGN (unresolved point source at the centre, $r<0\farcs06$ or $\approx$ 20 pc), the prominent starburst ring and bright stars. This MGE can be de-projected analytically with a specific axis ratio to reconstruct a 3D distribution of the entire galaxy. This best-fitting MGE model is tabulated in Table~\ref{tab_f547mmge} and shown in Fig.~\ref{hstF547mmge}, which illustrates the agreement between the data and the model in the form of two dimensional (2D) contours at the same radii and contour levels. 

{\it ISM mass:} \citet{Izumi20} estimated the total molecular mass of $M_{\rm H_2}\approx2.7\times10^9$~\Msun\ within the region of $r\leq3\arcsec$ ($\lesssim1$ kpc) using the dynamics of $^{12}$CO(2-1) and [CI](1-0). In this work, we estimated this mass using the $^{12}$CO(1-0) flux calculated in circular apertures. We converted these aperture fluxes into $M_{\rm H2}$ by assuming the CO-to-H$_2$ conversion factor for starburst galaxies: $X_{\rm CO}=1.0\pm0.3\times10^{20}$ cm$^{-2}$ (K~\kms)$^{-1}$ \citep{Kuno00, Kuno07, Bolatto13}.  We found the same amount of $M_{\rm H_2}$ within $3\arcsec$ and $M_{\rm H_2}(r\lesssim7\arcsec)\approx4\times10^9$~\Msun.

%%%%%%%%%%%%%%%%%%%%%%%%%%%%%%%%%%%%%%%%% 
\begin{figure*}
    \includegraphics[scale=0.65]{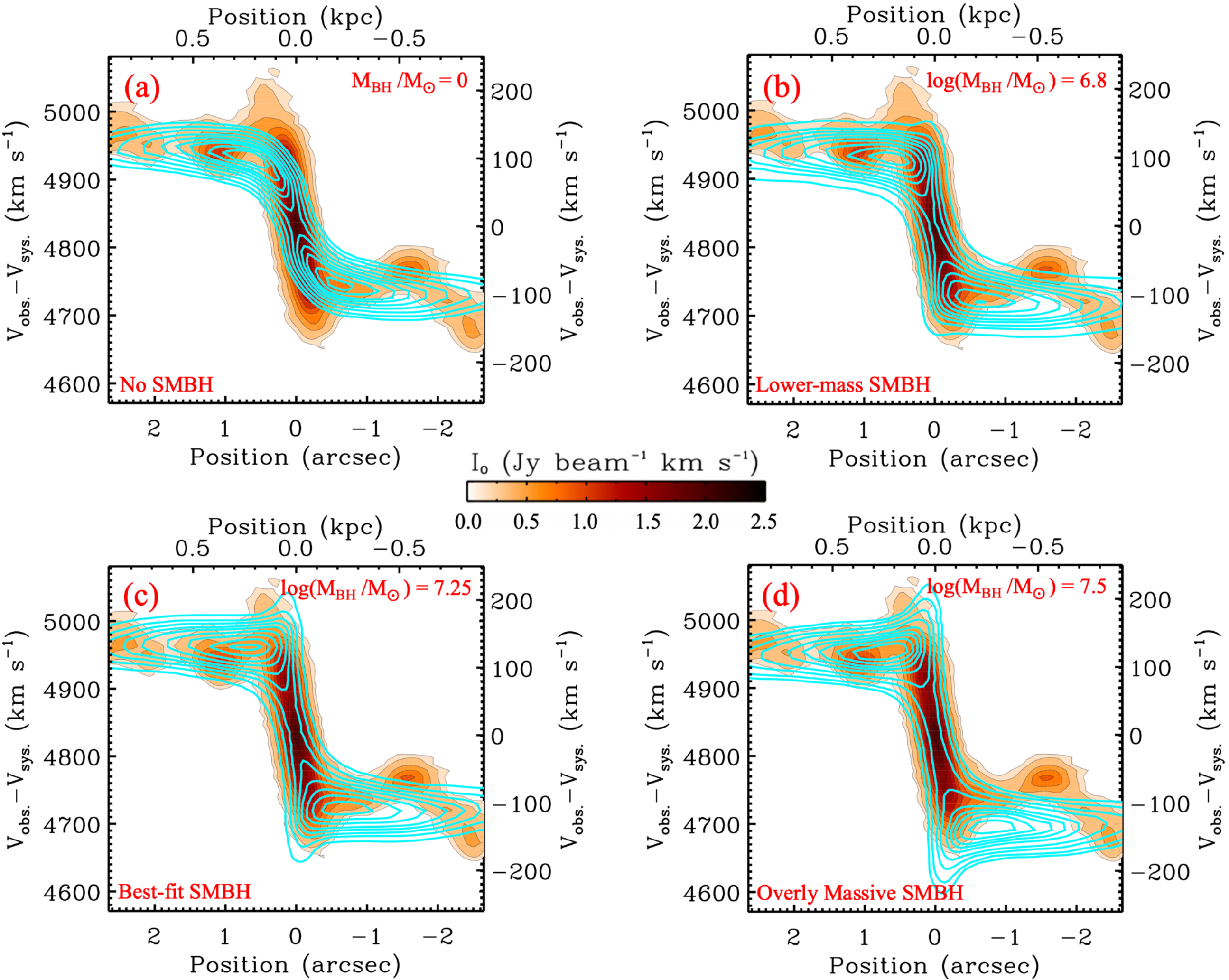} 
    \caption{Comparisons between data and a few KinMS models using the {\it default-mass model} and the [CI](1-0) emission. The PVD was extracted along the major-axis (orange scale and grey contours) in the same manner in Fig.~\ref{almapvd}. The model PVDs are extracted identically from models that are different in \Mbh\ (cyan contours), including the model without a SMBH (panel a) and the models with a small (panel b), the best fit (panel c, \Mbh\ =~$1.78\times10^7$~\Msun~and \ml$_{\rm F547M}=2.20$ (\Msun/\Lsun)), and an overly large (panel d) SMBH. The models in panels a, b, and d are not good fits for the data in the central part of the atomic-[CI](1-0) CND's kinematics as they fail to produce the raising rotation of the Keplerian motion towards the galaxy centre.}
    \label{bestfit_pvd2}   
\end{figure*}  
%%%%%%%%%%%%%%%%%%%%%%%%%%%%%%%%%%%%%%%%%

%%%%%%%%%%%%%%%%%%%%%%%%%%%%%%%%%%%%%%%%%
\begin{figure*}
    \centering\includegraphics[scale=0.65]{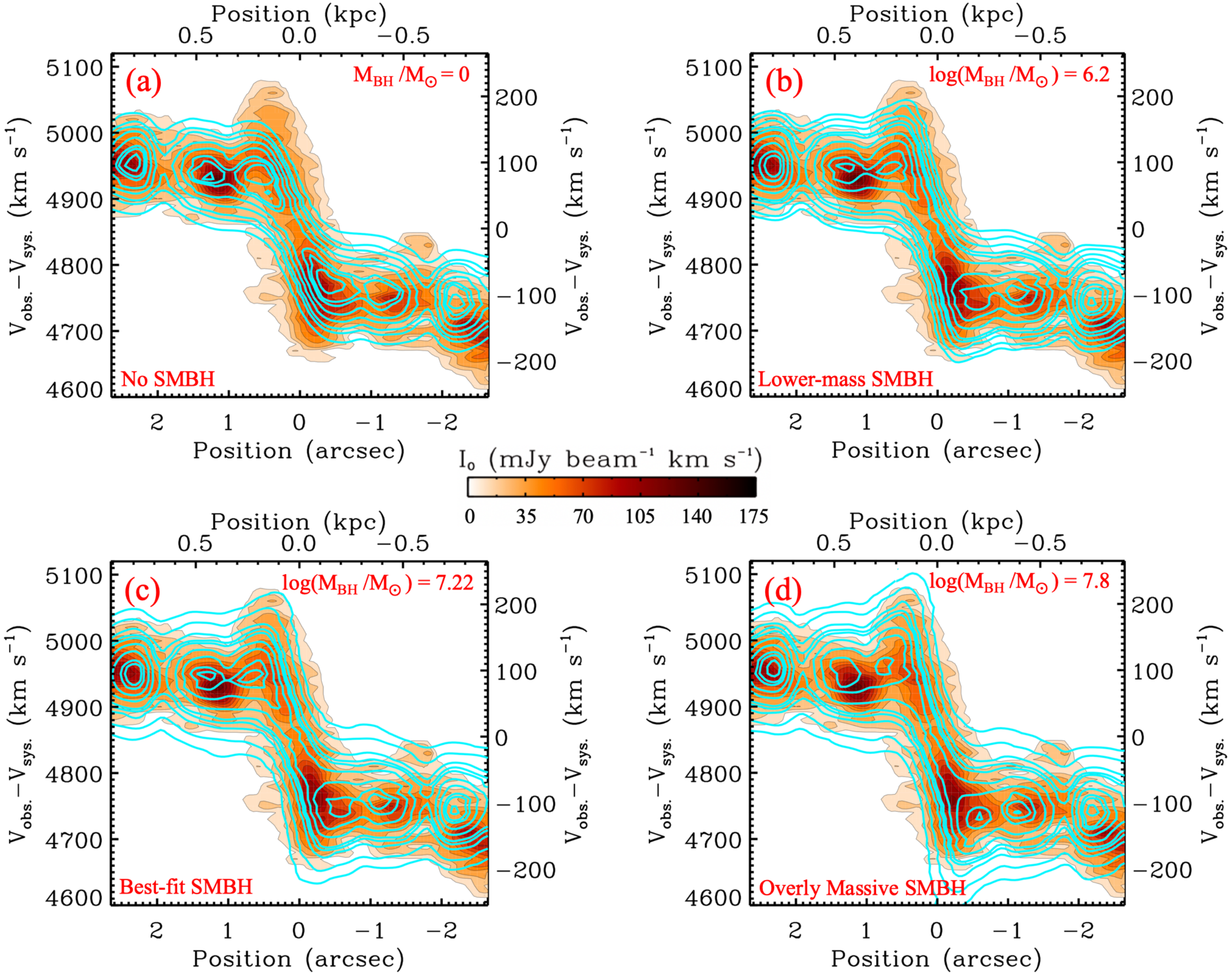} 
    \caption{Same as Fig.~\ref{bestfit_pvd2} but for $^{12}$CO(1-0). The best-fit model has \Mbh\ =~$1.6\times10^7$~\Msun\ and $M/L_{\rm F547M}=2.22$ (\Msun/\Lsun).}
    \label{bestfit_pvd1}   
\end{figure*}  
%%%%%%%%%%%%%%%%%%%%%%%%%%%%%%%%%%%%%%%%%

\citet{Papadopoulos00} also estimated the dust mass of $\approx2.5\times10^7$~\Msun\ in the same region of $r\leq7\arcsec$ ($\lesssim2.3$ kpc). These values give a gas-to-dust ratio of $\approx160$, which is close to the Galactic value. We then accounted for these ISM mass (molecular gas + dust) in the mass model by turning on the \texttt{gasgrav} mechanism in the KinMS model. Here, we assumed the molecular gas and dust are distributed accordingly to the surface brightness densities of $^{12}$CO(1-0) and [CI](1-0).

The stars and ISM masses directly predict the circular velocity of the $^{12}$CO(1-0) and [CI](1-0) CNDs using a constant \ml$_{\rm F547M}$ for the stellar-mass component as a free parameter in our dynamical models. We define the total mass model with this stellar-mass model as the {\it default-mass model} and quote its results throughout the paper.

%%%%%%%%%%%%%%%%%%%%%%%%%%%%%%%%%%%%%%%%%%%%%%%%%%%%%
\subsection{Results of the default-mass model}\label{defaultresult}

We performed the KinMS fit in an area of 60 pixels $\times$ 60 pixels ($6\arcsec\times6\arcsec\approx2\times2$ kpc$^2$). The model started with an initial guess of parameters that are flat priors in linear scales, except for \Mbh\ was in log-scale (to ensure efficient sampling of the posterior). The search ranges of these parameters are shown in column 2 of Table \ref{fittable}.  We limit the velocity channels ($-$300, 300) \kms\ related to the systemic velocity of $\approx4850$ \kms\ and $\approx4831$ \kms\ for $^{12}$CO(1-0) and [CI](1-0), respectively. To ensure our fit converges, we ran the model with $3\times10^5$ iterations, then exclude the first 20\% of the calculations as the burn-in phase and used the rest 80\% to produce the final posterior PDF of all nine free parameters.

Previous works showed that the ALMA noise covariance of a large number of high S/N-spaxels causes underestimates of the uncertainties, meaning the contribution of systematic errors dominates over the statistical errors \citep{Mitzkus17}. \citet{vandenBosch09} suggested that both errors give the same contribution and increased the $\Delta\chi^2$ required to define a given confidence level (CL) by the standard deviation of the $\chi^2$, namely $\sqrt{2(N-P)}\approx\sqrt{2N}$ \citep[see Section 15.1 of][]{Press07}, where $N=60\times60\times60$ is the number of constraints and $P=9$ is the number of model parameters or degree of freedom. Accordingly, in the Bayesian framework, we rescaled the model uncertainties by dividing the model $\log$ likelihood by $\sqrt{2N}$ or equivalently multiplying the {\tt rms} by $(2N)^{1/4}$ (Table \ref{tab_alma}). This rescaling approach was adopted in recent papers using KinMS and ALMA data \citep[][Nguyen et al. submitted; Smith al. submitted]{Onishi17, Nagai19, North19, Smith19, Davis20, Nguyen20} to yield more realistic uncertainties.

We interpreted our KinMS models' results with a SMBH at the centre of NGC~7469 to match the Keplerian motions of the $^{12}$CO(1-0) and [CI](1-0) kinematics. The best-fit model is identified via the Bayesian analysis of the post-burn-in PDFs generated by MCMC. Here, the best-fit value of each parameter is the median of the marginalised parameter posterior PDFs of all other parameters, and its uncertainty are all models within (31--69\%), (16--84\%), (2.3--97.7\%), and (0.14--99.86\%) of the PDFs or 0.5$\sigma$, 1$\sigma$, 2$\sigma$, and 3$\sigma$ CLs, respectively.

{\it \Mbh\ derived with the atomic-[CI](1-0) kinematics:} At $3\sigma$ CL, the best-fit model gives $M_{\rm BH}=1.78^{+2.69}_{-1.10}\times10^7$~\Msun\ and $M/L_{\rm F547M}=2.20^{+0.43}_{-0.40}$ (\Msun/\Lsun) at $\chi^2_{\rm min,\;red}=1.94$. Other best-fit parameters and their likelihoods show in Table \ref{fittable}. 

{\it \Mbh\ derived with the molecular-$^{12}$CO(1-0) kinematics:} Similarly, at $3\sigma$ CL, This best-fit model has $M_{\rm BH}=1.60^{+11.52}_{-1.45}\times10^7$~\Msun\ and $M/L_{\rm F547M}=2.22^{+0.20}_{-0.22}$ (\Msun/\Lsun) at $\chi^2_{\rm min,\;red}=1.91$. Also, see Table \ref{fittable} for the full model's description.

\begin{figure*}
   \includegraphics[scale=0.58]{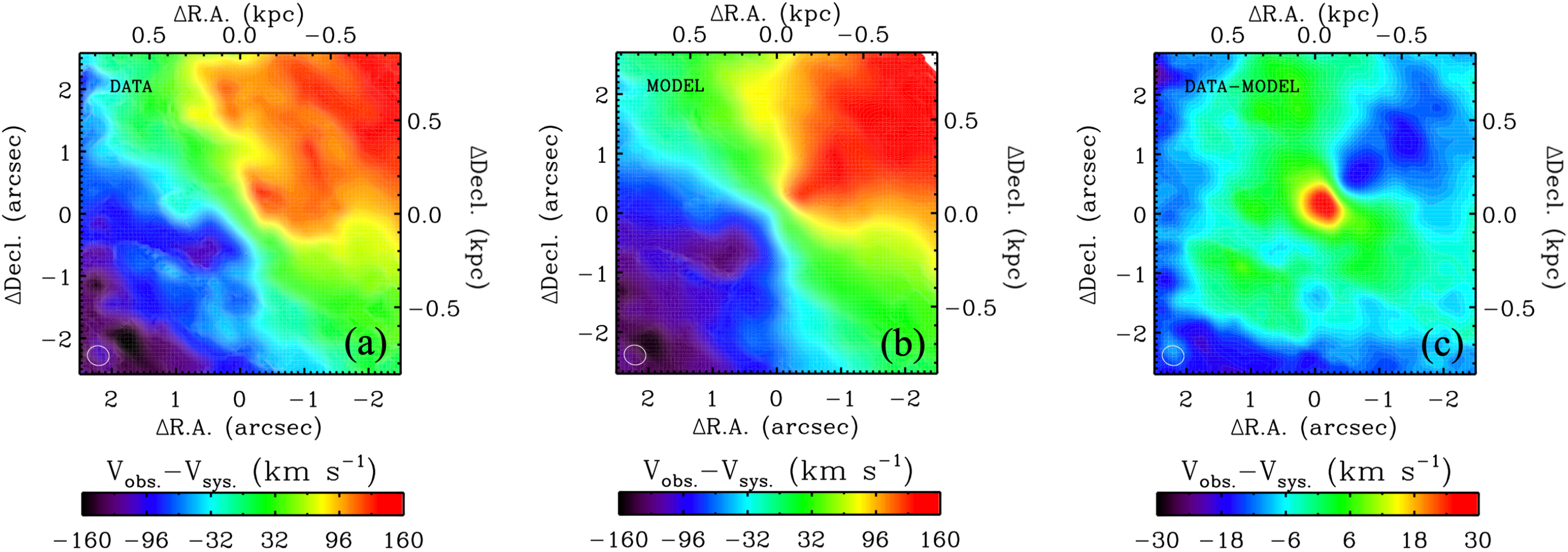} 
    \caption{The $4.8\arcsec\times4.8\arcsec$ ($\approx1.7\times1.7$ kpc$^2$) FOV of the intensity-weighted mean LOS velocity field map of [CI](1-0) (panel a), the map derived from the best-fit KinMS model shown in panel c of Fig.~\ref{bestfit_pvd2} (panel b), and the residual {\tt (Data - Model)} LOS map (panel c).}
    \label{bestfit_mom22}   
\end{figure*}

\begin{figure*}
    \includegraphics[scale=0.58]{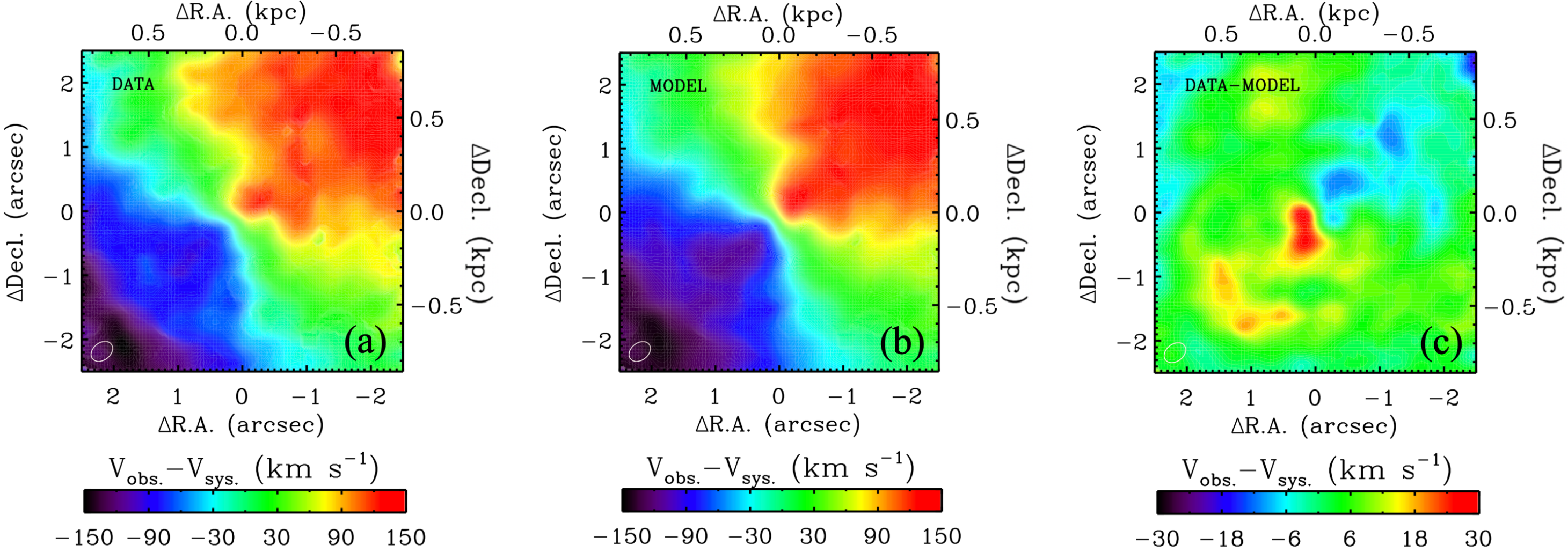} 
    \caption{Same as Fig.~\ref{bestfit_mom22} but for $^{12}$CO(1-0). The best-fit KinMS model is the one shown in panel c of Fig.~\ref{bestfit_pvd1}.}
    \label{bestfit_mom11} 
\end{figure*}

To demonstrate how well the models describe the observables, we show the observed PVDs overlaid with the best-fit PVDs in Fig.~\ref{bestfit_pvd2} for [CI](1-0) and Fig.~\ref{bestfit_pvd1} for $^{12}$CO(1-0). The models without, with lower-mass, and with overly massive SMBHs also add for comparisons. All these models do not fit the CND's central Keplerian motions. The intensity-weighted mean LOS velocity field of the data, the best-fit models, and their residual fields ${\tt (Data-Model)}$ within the FOV of $5\farcs2\times5\farcs2$ ($\approx1.7\times1.7$ kpc$^2$) show in Fig.~\ref{bestfit_mom22} for [CI](1-0) and Fig.~\ref{bestfit_mom11} for $^{12}$CO(1-0). Within this FOV, most of the residual map has the amplitude of $\approx10$~\kms, the limit in which we bin the spectrum together (the channel width or the uncertainty of our kinematic measurements). However, there is a large residual ($\approx30$~\kms) on the northwest and southeast sides of the galaxy centre for [CI](1-0) and $^{12}$CO(1-0), respectively, which might be interpreted as real non-circular motions. Nevertheless, these high-velocity residuals in both emissions are perhaps the results of the asymmetric rotation seen in panels d of Figs~\ref{momsmap1} and \ref{momsmap2} (Section~\ref{almaobs}), or perhaps due to the beam smearing effects because of the limited-spatial resolutions of our data, and consequently, the black hole could be a bit more massive. Future higher resolution observations will better estimate these non-circular motions, which will deliver a more robust \Mbh.

\begin{figure*}
    \centering\includegraphics[scale=0.34]{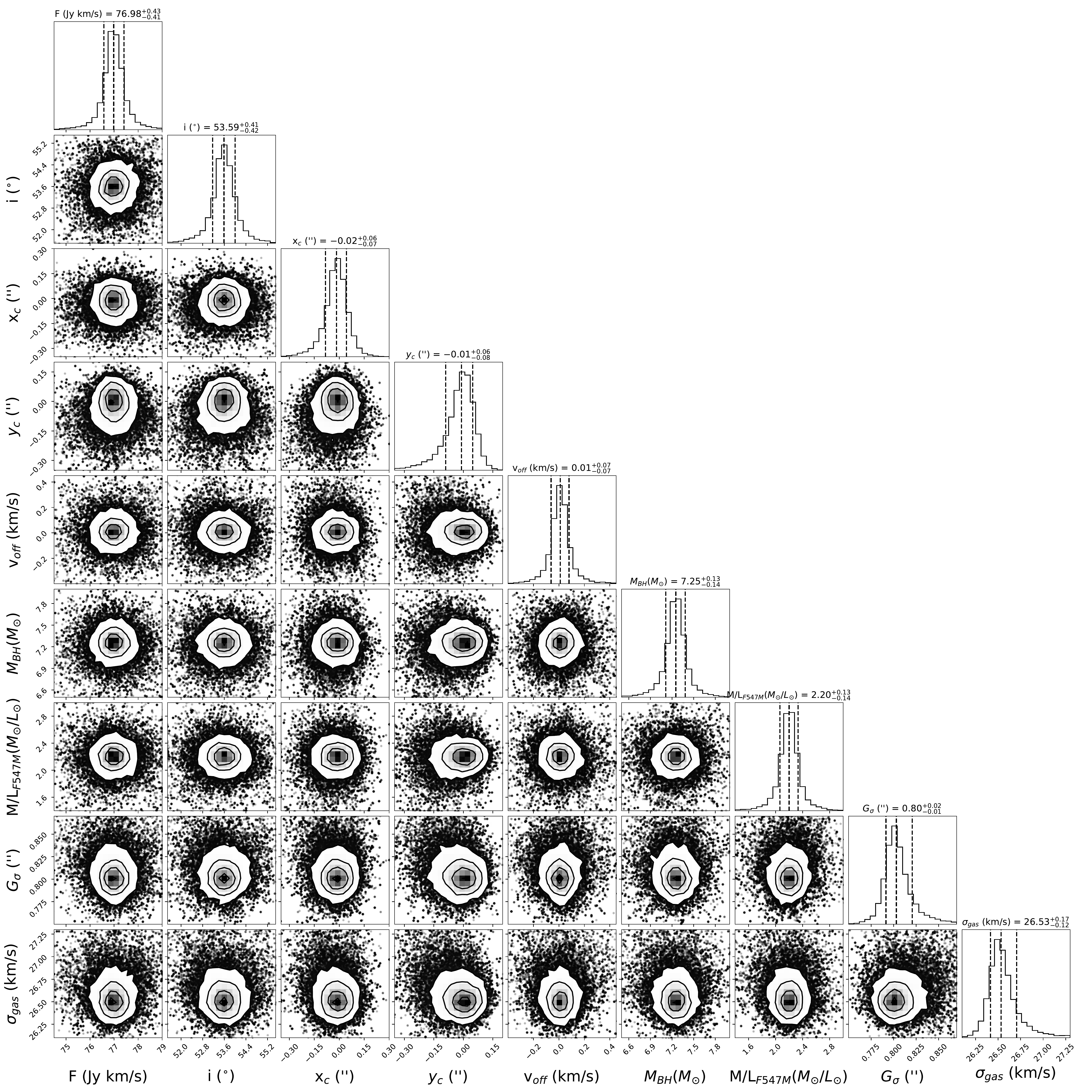} 
    \caption{The posterior distribution of the KinMS model's parameters, which use the {\it default-mass} model and the [CI](1-0) kinematics, explored in the Bayesian framework using the MCMC technique to fit the central $5\arcsec\times5\arcsec$ ($\approx1.7\times1.7$ kpc$^2$) FOV kinematics. The top panel of each column is the marginalised PDF of the associated parameter, with the median and 16--84\% (i.e. 1$\sigma$). The lower panels show the 2D marginalisations of each model parameter. Out from the centre, the thick solid contours indicate the 0.5$\sigma$ (31--69\%), 1$\sigma$ (16--84\%), 2$\sigma$ (2.3--97.7\%), and 3$\sigma$ (0.14--99.86\%) CLs. See Table \ref{fittable} for a quantitative description of the likelihoods of all fitting parameters. \Mbh\ is presented in log scale. For illustrations of how well this model describes the data, readers can refer to Fig.~\ref{bestfit_pvd2}.}  
\label{posteriorci}   
\end{figure*}

\begin{figure*} 
    \centering\includegraphics[scale=0.34]{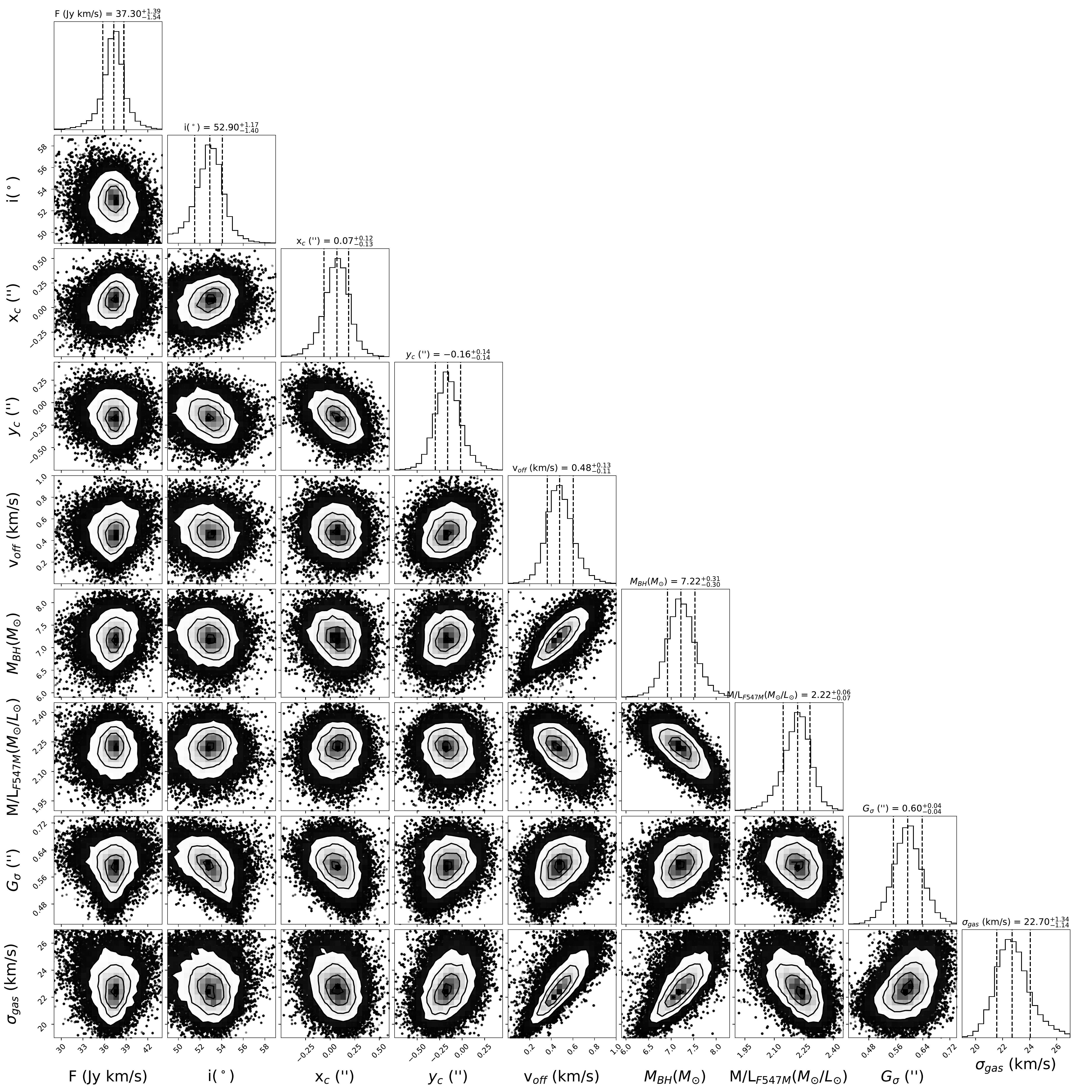} 
    \caption{Same as Fig.~\ref{posteriorci} but for $^{12}$CO(1-0). \Mbh\ is presented in log scale. Readers can refer to Figs~\ref{bestfit_pvd1} and \ref{bestfit_mom11} for how well this model describe the data.}
 \label{posteriorco}   
\end{figure*}

We show the PDF and 2D marginalisations of each free parameters of KinMS models of [CI](1-0) and $^{12}$CO(1-0) in Figs \ref{posteriorci} and \ref{posteriorco}, respectively. There is no covariance found among parameters of the [CI](1-0) kinematics. For $^{12}$CO(1-0), covariance of \Mbh\ vs. \ml$_{\rm \rm F547M}$ is present as the result of the degeneracy between the potentials of the SMBH and galaxy itself, happening when the observational scales have not resolved the SMBH's SOI.  Also, other covariances of $G_\sigma$ and $\sigma_{\rm gas}$ vs. other parameters in the two bottom row panels of its posterior cause by no clear central peak in the $^{12}$CO(1-0) integrated intensity map (panel a of Fig.~\ref{momsmap1}). Three bright clumps of high-brightness-density at the $^{12}$CO(1-0) CND's centre increases the explored ranges for $x_{\rm c}$ and $y_{\rm c}$, then resulted in such covariances among the parameters during a simultaneous constraint. 

As we mentioned that the atomic-[CI](1-0) line might be a better transition to do dynamical modelling for the central  \Mbh\ than the molecular-$^{12}$CO(1-0) line in Section~\ref{sec:intro}. {\it We proved here that the KinMS model using the [CI](1-0) kinematics produces a significantly smaller \Mbh-uncertainty than that of the KinMs model using the $^{12}$CO(1-0) kinematics}. Also, there is no correlation between the constrained- \Mbh\ and \ml$_{\rm F547M}$ (Fig.~\ref{posteriorci}), meaning an excellent constraint. We thus quote the parameter values inferred from the KinMS model using the kinematics measured from the atomic-[CI](1-0) line throughout this work.  However, it should be noticed that although the \ml$_{\rm F547M}$-uncertainty inferred from the KinMS model using the $^{12}$CO(1-0) kinematics is a factor of $\approx2$ smaller (in log scale) than that of the KinMS model using the [CI](1-0) kinematics, the correlation between its constrained- \Mbh\ and \ml$_{\rm F547M}$ is present (Fig.~\ref{posteriorco}). This \ml$_{\rm F547M}$ uncertainty mismatch is probably caused by (1) the more extended distribution of $^{12}$CO(1-0) kinematics ($\approx7\arcsec$) comparing to that of [CI](1-0) ($<5\arcsec$) as seen in panels a and b of Fig.~\ref{hstalma}, meaning it is the better probe of the extended mass or \ml$_{\rm F547M}$. Also, (2) the somewhat higher symmetry of the extended kinematics of $^{12}$CO(1-0) than that of [CI](1-0) seen in their extracted PVDs along the galaxy-major axis (the lower-velocity wings Fig.~\ref{almapvd}) help to reduce its \ml$_{\rm F547M}$ uncertainty (see the consistencies between the data and the best-fit models in panels c of Figs~\ref{bestfit_pvd2} and \ref{bestfit_pvd1}). 

\citet{Boizelle19} discussed the angular-resolution limit at which one can constrain a reliable \Mbh\ is $\theta_{\rm beam}\lesssim\theta_{\rm r_{SOI}}$, where $\theta_{\rm r_{SOI}}$ is the angle subtended by $r_{\rm SOI}$. Measurements using data with larger synthesised beams are more susceptible to systematic biases from stellar-mass uncertainties. Given $M_{\rm BH}\approx1.8\times10^7$~\Msun\ and $\sigma_\star\approx150$~\kms\ \citep{Onken04} gives $r_{\rm SOI}\equiv GM_{\rm BH}/\sigma_\star^2\approx3$ pc ($\approx0\farcs01$), a factor of $\approx30$ times smaller than what our ALMA observations can resolve. This work thus belongs to the majority of \Mbh\ measurements that have $\theta_{\rm beam}>\theta_{\rm r_{SOI}}$ \citep[][Nguyen et al. submitted]{Davis13, Davis17, Davis18, Onishi15, Onishi17, Smith19, Nguyen20} and the lowest angular resolution with \Mbh\ ever measured. Despite such a large $\theta_{\rm beam}$, our dynamical models using the atomic-[CI] kinematics can constrain a robust \Mbh\ within a robust systemic/statistic uncertainty \citep[$\gtrsim35$\% when $\theta_{\rm beam}>\theta_{\rm r_{\rm SOI}}$;][]{Nguyen20}. While the same constraints using the molecular-$^{12}$CO(1-0) emission were subject to much larger errors with possible estimated-\Mbh\ that are all within the mass range of $10^6-10^8$~\Msun\ (Table~\ref{fittable}). The main reason is that the atomic-[CI] emission is highly concentrated at the galaxy centre (close to the SMBH) due to XDR effects around the AGN (I20), while the molecular-$^{12}$CO(1-0) emission is much more extended (Fig.~\ref{hstalma}). However, observations at higher angular resolutions would benefit our \Mbh\ estimate by further reducing the uncertainties arising from the stellar mass.

It is also worth mentioning that the dynamical modeling of the other CO lines observed in this program did not improve the inferred-\Mbh\ uncertainty significantly comparing to the chosen $^{12}$CO(1-0). These molecular emissions distributions are a bit more compact towards the centre of NGC~7469, but their spatial resolutions are a little lower than $^{12}$CO(1-0). Additionally, the appearance of the $^{12}$CO(2-1) and $^{12}$CO(3-2) PVDs are similar to that of $^{12}$CO(1-0) PVD. i.e., there would be a ``hole'' at the AGN position \citep[or CO-deficit;][]{Smith19, Nguyen20}, hinted by three resolved clumps in the integrated intensity map of the $^{12}$CO(1-0) emission in panel a of Fig.~\ref{momsmap1}; also see Fig.~2 of \citet{Izumi20}. Furthermore, the compactness of the molecular emission has an advantage in reducing the \Mbh\ uncertainty but increasing the \ml$_{\rm F547M}$ uncertainty in the same manner as seen in the case of [CI](1-0), which is clearly centrally-peaked (panel a of Fig.~\ref{momsmap2}).

%%%%%%%%%%%%%%%%%%%%%%%%%%%%%%%%%%%%%%%%%%%%%%%%%%%%%%%%%%%
\subsection{The stellar-mass models account for spatial variations of populations and extinction}\label{colourmass}

\begin{figure}
    \centering\includegraphics[scale=0.4]{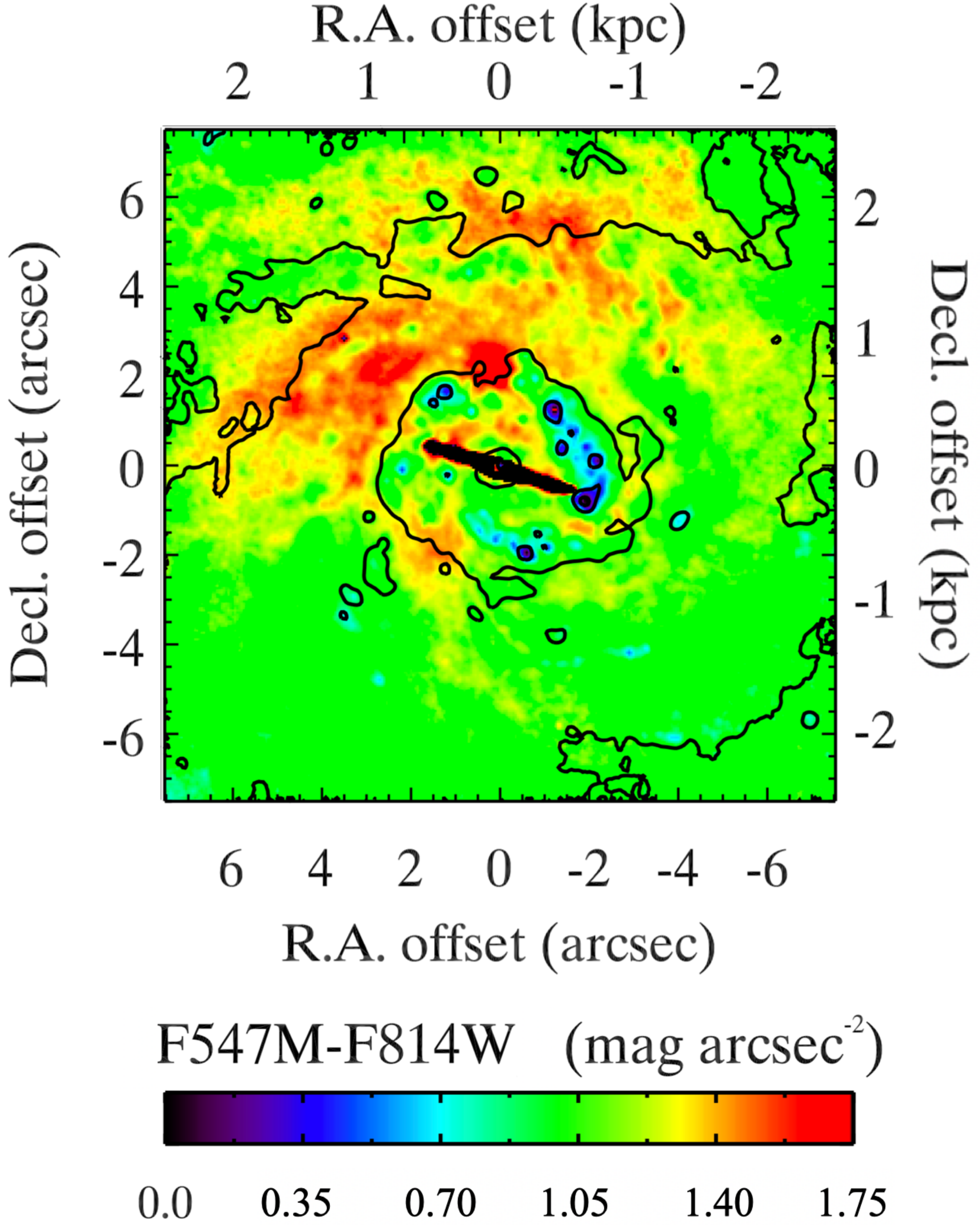} 
    \caption{The \hst~F547M--F814W ($\approx$ $V-I$) colour map within $7\farcs5$ ($\approx2.5$ kpc) nuclear region. The map centres at the ALMA continuum peak. Redder areas are resulted from older populations and dust extinction, while bluer areas are resulted from very young star-forming regions. The southern and western half of the nucleus appears to have little extinction. The black trail at the centre includes the excluded spaxels of saturation in the F814W image. The black contours show the \hst/F547M continuum emission at the surface brightness density of 17.2, 18.3, and 19.1 mag arcsec$^{-2}$ \citep[also see the photometry in $V$-band of][]{Doroshenko98}.}
    \label{colourmap}   
\end{figure}

There is a colour variation due to complex stellar populations and dust extinction in the nucleus region of  NGC~7469 shown in Fig.~\ref{colourmap}, implying that our assumption of a constant \ml\ is too simple. Here, we used an additional \hst/F814W ($\approx I$) image (Table \ref{tab_hst}) in combination with the F547M ($\approx V$) image to create a F547M--F814W colour map (Fig.~\ref{colourmap}). Next, we convolved each astrometrically aligned image to the other PSF image (e.g. the F547M image convolved with the F814W PSF) to mitigate spurious gradients near the centre of the galaxy due to the difference of PSF widths. Then, we subtracted off the sky background level on each image calculated in an annulus located at (20--22)$\arcsec$ away from the galaxy centre.

\begin{figure*}
    \centering\includegraphics[scale=0.55]{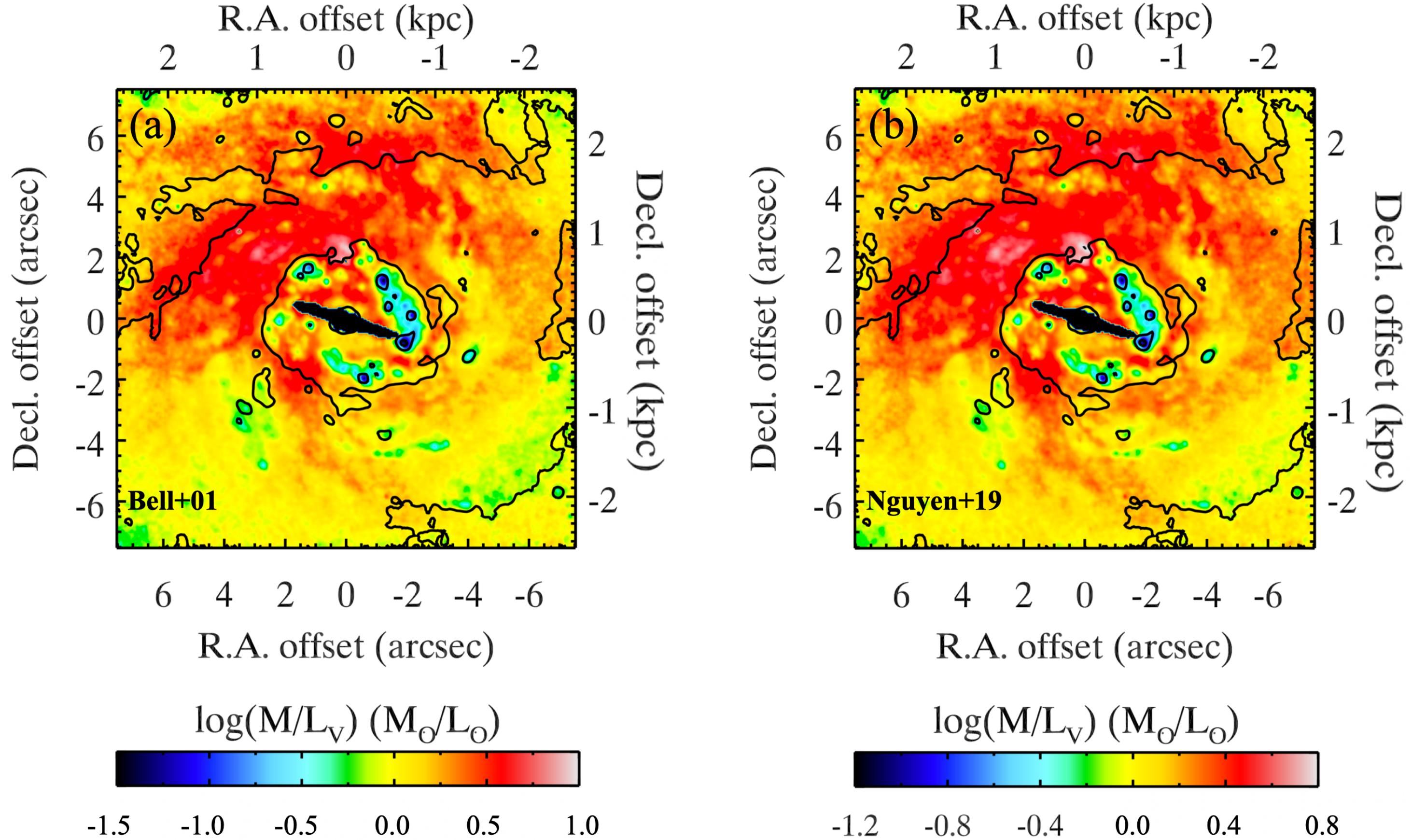}\vspace{5mm}
   \includegraphics[scale=0.55]{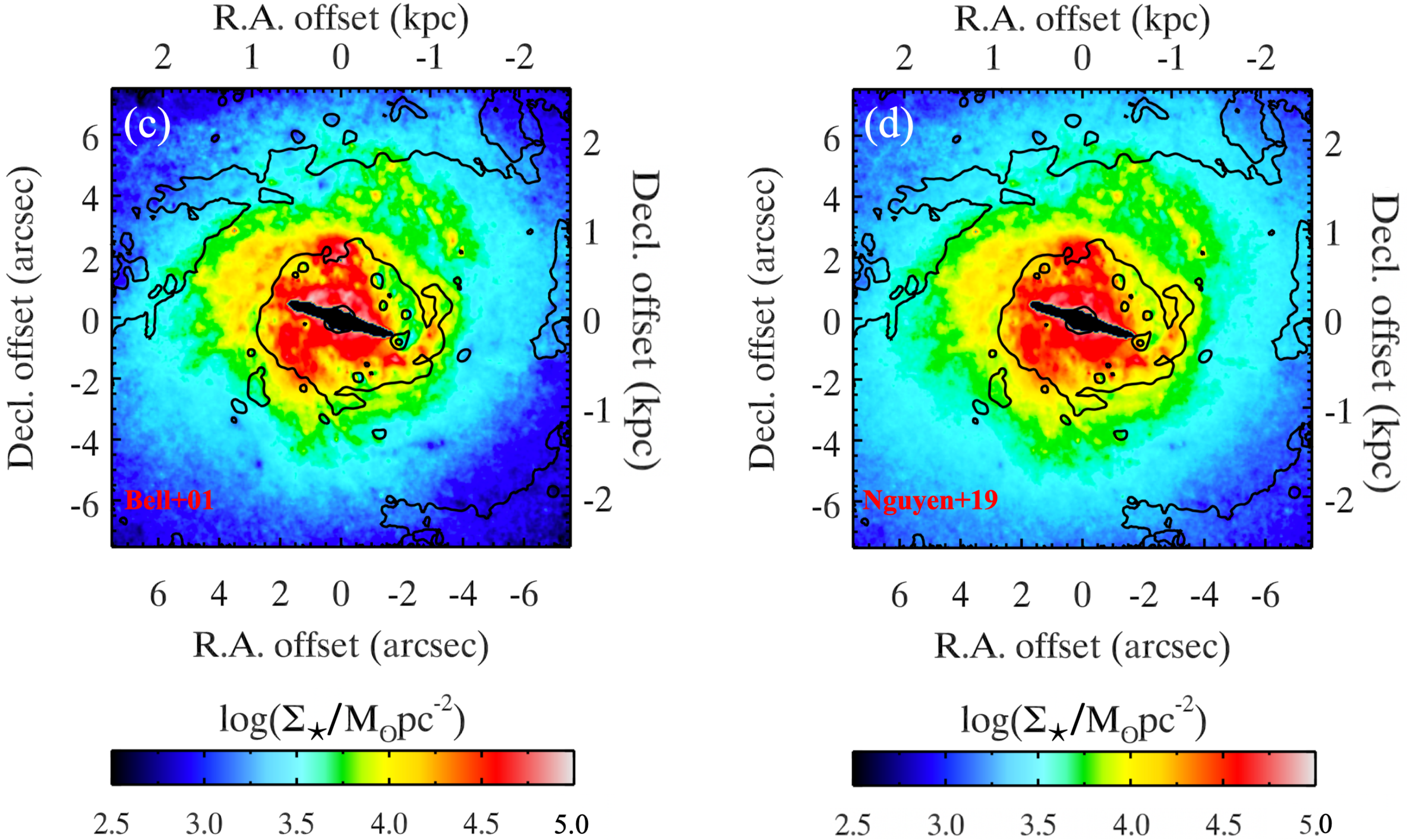} 
    \caption{The \ml$_V$ maps constructed from the F547M--F814W colour map using the B01 (panel a) and the N19 (panel b) colour--\ml~correlation. Panel c: The nuclear surface-density-mass maps of NGC~7469 constructed by multiplying the F547M surface-density-luminosity map to the B01 (panel c) and the N19 (panel d) \ml$_V$ map pixel by pixel. Other illustrations (e.g. contours and saturated spaxels) are similar to Fig.~\ref{colourmap}.}
    \label{m2lmap}   
\end{figure*}

There is a heavy dust-extinction on the northern and northeast sides, while the southern and southwest sides of the nucleus have little extinction. Also, the starburst ring appears to host very young star clusters, suggesting the \ml\ variation. Such variation had a significant impact on the dynamical models of $\lesssim$$10^7$~\Msun\ SMBHs \citep[][N19]{McConnell13b, Thater19a, Nguyen18}. In principle, we could apply the colour--\ml\ scaling relations for the young nucleus ($<$1 Gyr) compiled by B01 and N19 to convert this colour map into their corresponding \ml\ maps (panels a and b of Fig.~\ref{m2lmap}) and mass-surface-density maps (panels c and d of Fig.~\ref{m2lmap}) that account for the variation of stellar populations pixel by pixel. However, the central saturation found in F814W prevents us from creating an accurate mass-surface density at the galaxy centre, where the SMBH's gravitational potential dominates.

The N19-mass map predicts less mass at redder regions of contaminated dust and more mass at the bluer regions of young stars in the starburst ring than the B01-mass map's prediction. It is also much more symmetric and thus is a better description of true mass-distribution in the nucleus of NGC~7469 because the N19 colour--\ml\ correlation specifically applied for young populations ($<$1 Gyr) and accounted for the extinction by dust. On the other hand, the B01 relation only relies on the spectrophotometric evolution models of spiral galaxies and the colours of the integrated stellar populations. 

We converted the B01 and N19 mass-surface-density maps into their MGE forms with the central stellar mass in the centrally saturated spaxels interpolated from the outer pixels. Then, in their KinMS models, we first replaced the \ml$_{\rm F547M}$ parameter by the mass-scaling factor $\Gamma=(M/L_{\rm dyn})/(M/L_{\rm pop})$ to scale the mass-surface-density profiles directly, where $(M/L_{\rm dyn})$ is the mass-to-light ratio constrained from dynamical model and $(M/L_{\rm pop})$ is that determined from stellar populations. Next, we allowed \Mbh\ and $\Gamma$ to vary as free parameters, fixed other well-constrained parameters at their best fits shown in Table~\ref{fittable}, and ran the model with the kinematics of the atomic-[CI](1-0) emission only. These resulted in $M_{\rm BH}=2.2^{+3.3}_{-1.3}\times10^7$~\Msun\ and $\Gamma=1.09^{+0.21}_{-0.23}$ for the B01-mass map and $M_{\rm BH}=2.9^{+1.8}_{-1.7}\times10^7$~\Msun\ and $\Gamma=0.93^{+0.20}_{-0.20}$ for the N19-mass map. Each of these masses is a factor of $\approx$ 1.3 and 1.6 higher than our best-fit model in Section~\ref{defaultresult}, indicating that the specific stellar populations and dust extinction in the nucleus of NGC~7469 have somewhat of an impact on our dynamical modelling. 

%%%%%%%%%%%%%%%%%%%%%%%%%%%%%%%%%%%%%%%%%
%%%%%%%%%%%%%%%%%%%%%%%%%%%%%%%%%%%%%%%%%

%%%%%%%%%%%%%%%%%%%%%%%%%%%%%%%%%%%%%%%%%
%%%%%%%%%%%%%%%%%%%%%%%%%%%%%%%%%%%%%%%%

%%%%%%%%%%%%%%%%%%%%%%%%%%%%%%%%%%%%%%%%%
%%%%%%%%%%%%%%%%%%%%%%%%%%%%%%%%%%%%%%%%%

%%%%%%%%%%%%%%%%%%%%%%%%%%%%%%%%%%%%%%%%%
%%%%%%%%%%%%%%%%%%%%%%%%%%%%%%%%%%%%%%%%%

%%%%%%%%%%%%%%%%%%%%%%%%%%%%%%%%%%%
\begin{table*}
\caption{Best-fitting model parameters and associated statistical uncertainties for the {\it default-mass model} and the kinematics of [CI](1-0) and $^{12}$CO(1-0)}
\begin{tabular}{lcccc||||||||||||ccc} 
\hline\hline       
Parameters (Units)                          & Search range&Best fit &  $1\sigma$ error   &    $3\sigma$ error   &Best fit & $1\sigma$ error  &  $3\sigma$ error      \\
                                                        &  (Uniform)  &       & (1--84\%)  & (0.14--99.86\%) &        & (1--84\%)  & (0.14--99.86\%)       \\ 
\hline 
{\it Emission Line}                           &              &        &[CI](1-0)&             &        &$^{12}$CO(1-0)&               \\
\hline 
{\it Black Hole:}                               &              &        &              &                 &        &            &                 \\
$\log(M_{\rm BH}/{\rm M_\odot})$  &    (1, 9)    &  7.25  &$-$0.14, +0.13&  $-$0.42, +0.40 &  7.22  &$-$0.30, +0.31&$-$0.90, +0.93 \\
$M/L_{\rm F547M}$ (\Msun/\Lsun)&(0.1, 5.0) &  2.20  &$-$0.14, +0.13&  $-$0.43, +0.40 &  2.22  &$-$0.07, +0.06&$-$0.22, +0.20\\
{\it Gas CNDs:}                        &              &        &              &                 &        &              &                       \\
$\sigma_{\rm gas}$ (\kms)            &   (1, 50)    & 26.53  &$-$0.12, +0.17&  $-$0.36, +0.40 &  22.70 &$-$1.14, +1.34&$-$4.00, +4.02\\
$G_\sigma$    ($\arcsec$)            &   (0.1, 1.5) &  0.80  &$-$0.01, +0.02&  $-$0.03, +0.06 &  0.60  &$-$0.04, +0.04&$-$0.12, +0.12\\ 
$F$           (Jy \kms)                      &   (1, 100)   &  76.98 &$-$0.41, +0.43&  $-$1.23, +1.29 &  37.30 &$-$1.54, +1.39&$-$4.51, +4.32\\
$i\;(^\circ)$                                    &   (45, 89)   &  53.59 &$-$0.42, +0.41&  $-$1.26, +1.25 &  52.90 &$-$1.40, +1.17&$-$3.20, +3.10\\
{\it Nuisance:}                               &              &        &              &                 &        &              &                        \\
$x_{\rm c}$        ($\arcsec$)        &($-$1.5, +1.5)&$-$0.02 &$-$0.07, +0.06&  $-$0.21, +0.20 & $+$0.07&$-$0.13, +0.12&$-$0.38, +0.36\\
$y_{\rm c}$        ($\arcsec$)        &($-$1.5, +1.5)&$-$0.01 &$-$0.08, +0.07&  $-$0.24, +0.21 & $-$0.16&$-$0.14, +0.14&$-$0.42, +0.42\\
$v_{\rm off}$   (\kms)                   &  ($-$5, +5)  & 0.01   &$-$0.07, +0.07&  $-$0.21, +0.21 &  0.48  &$-$0.11, +0.13&$-$0.33, +0.37\\
\hline
\end{tabular}
\parbox[t]{0.92\textwidth}{\textit{Notes: }{The table columns list respectively each parameter name, search range, best fit and uncertainty at $1\sigma$ (1--84\%) and $3\sigma$ (0.14--99.86\%) CLs of the PDF. The parameters $x_{\rm c}$, $y_{\rm c}$ and $v_{\rm off}$ are parameters defined relative to the adopted galaxy centre $(23^{\rm h}03^{\rm m}15\fs617$, $+08\degr52\arcmin26\farcs00$) and $V_{\rm sys.}=4831$~\kms\ and $4834$ \kms\ for [CI](1-0) and $^{12}$CO(1-0), respectively. }} 
\label{fittable}
\end{table*}
%%%%%%%%%%%%%%%%%%%%%%%%%%%%%%%%%%%

%%%%%%%%%%%%%%%%%%%%%%%%%%%%%%%%%%%%%%%%%
%%%%%%%%%%%%%%%%%%%%%%%%%%%%%%%%%%%%%%%%%

%%%%%%%%%%%%%%%%%%%%%%%%%%%%%%%%%%%%%%%%%%%
\subsection{Other uncertainty sources on the \Mbh\ estimate}\label{uncertainty}

We test here the robustness of our dynamical model under the influence of sources of error other than the uncertainties in ALMA kinematics and stellar-mass errors. To this end, we tested dynamical models with the atomic-[CI](1-0) kinematics and allowed only the \Mbh\ and \ml$_{\rm F547M}$ changing as free parameters while fixed other parameters at their best-fitting values in Table~\ref{fittable}. The results of these tests include the best-fitting parameters associated with 1$\sigma$ (16--84\%) and 3$\sigma$ (0.14--99.86\%) CLs recorded in Table~\ref{tab:bhtest}.

%%%%%%%%%%%%%%%%%%%%%%%%%%%%%%%%%%%%%%%%%%%%
\subsubsection{\citet{Izumi20} CO-to- vs [CI]-to-H$_2$ conversion factor}\label{conversion}

Our choice of the starbursting galaxy conversion factor to estimate the molecular gas mass in Section~\ref{massmodels} is uncertain because it could vary on small scales of kinematics and could also be lower in the very centre due to stellar processes. \citet{Meier08} have found that $X_{\rm CO}$ varies in the range of $(0.5-1)\times10^{20}$ cm$^{-2}$ (K \kms)$^{-1}$ in local normal spiral galaxies due to spiral arm/bar streaming. It is also too high in clouds with substantial stellar-content. Based on dynamical modellings, I20 found the CO-to- and [CI]-to-H$_2$ conversion factor of $X_{\rm CO}=1.9\times10^{20}$ cm$^{-2}$ (K \kms)$^{-1}$ and $X_{\rm [CI](1-0)}=2.1\times10^{20}$ cm$^{-2}$ (K \kms)$^{-1}$, respectively, at the innermost  $\approx100$ pc region of NGC~7469. The CO-to-H$_2$ conversion factor is, coincidentally equal to, while the [CI]-to-H$_2$ conversion factor is smaller than, those derived for Galactic star-forming regions due to the elevated C$^0$ abundance in the XDR. These suggest that our original adoption of the starburst CO-to-H$_2$ conversion factor may be inappropriate and may underestimate the total molecular gas mass by a factor of $\approx2$. We, therefore, used instead these newly derived conversion factors of NGC~7469 to test our modellings. The best-fit KinMS models then yield $M_{\rm BH}=9.3_{-4.6}^{+9.2}\times10^6$~\Msun\ and \ml$_{\rm F547M}=2.27_{-0.45}^{+0.46}$~(\Msun/\Lsun) for CO-to-H$_2$ and $M_{\rm BH}=9.8_{-5.5}^{+12.6}\times10^6$~\Msun\ and \ml$_{\rm F547M}=2.30_{-0.43}^{+0.43}$~(\Msun/\Lsun) for [CI]-to-H$_2$, suggesting these adopted conversion factors have a significant impact on the derived \Mbh\ but within $3\sigma$ uncertainties of our best-fit models in Section~\ref{defaultresult} and Table~\ref{fittable}).

%%%%%%%%%%%%%%%%%%%%%%%%%%%%%%%%%%%%%%%%%
%%%%%%%%%%%%%%%%%%%%%%%%%%%%%%%%%%%%%%%%%

%%%%%%%%% Table: Best-fitting model parameters %%%%%%%%%%%%%%%%%%%%%
\begin{table}
  \caption{Best-fitting model parameters and associated statistical uncertainties of the atomic-[CI](1-0) kinematics.}
  \begin{tabular}{lcccccr} 
    \hline\hline       
    Parameter & Best fit & $1\sigma$ uncertainty &  $3\sigma$ uncertainty \\
              &  & ($16$--$84\%$) & ($0.14$--$99.86\%$) \\  
    (1) & (2) & (3) & (4) \\  
    \hline
    \multicolumn{4}{l}{\bf \citet{Bell01} mass model} \\  
    \hline
    $\log(M_{\rm BH}/{\rm M}_\odot)$ & $\phantom{7}7.35$ & $-0.13$, $+0.13$ & $-0.40$, $+0.38$\\
    $\Gamma$                       & $\phantom{7}1.09$ & $-0.08$, $+0.07$ & $-0.23$, $+0.21$\\
    \hline
    \multicolumn{4}{l}{\bf \citet{Nguyen19a} mass model} \\  
    \hline
    $\log(M_{\rm BH}/{\rm M}_\odot)$ & $\phantom{7}7.46$ & $-0.14$, $+0.13$ & $-0.42$, $+0.40$\\
    $\Gamma$                       & $\phantom{7}0.93$ & $-0.07$, $+0.07$ & $-0.21$, $+0.21$\\
    \hline 
    \multicolumn{4}{l}{\bf \citet{Izumi20} CO-to-H$_2$ conversion factor} \\  
    \hline
    $\log(M_{\rm BH}/{\rm M}_\odot)$ & $\phantom{7}6.97$ & $-0.10$, $+0.10$ & $-0.30$, $+0.30$\\
    \ml$_{\rm F547M}$ (\Msun/\Lsun) & $\phantom{7}2.27$ & $-0.15$, $+0.15$ & $-0.45$, $+0.46$\\
    \hline 
    \multicolumn{4}{l}{\bf \citet{Izumi20} [CI]-to-H$_2$ conversion factor} \\  
    \hline
    $\log(M_{\rm BH}/{\rm M}_\odot)$ & $\phantom{7}6.99$ & $-0.12$, $+0.12$ & $-0.36$, $+0.36$\\
    \ml$_{\rm F547M}$ (\Msun/\Lsun) & $\phantom{7}2.30$ & $-0.14$, $+0.14$ & $-0.43$, $+0.43$\\
    \hline
    \multicolumn{4}{l}{\bf Axisymmetric ISM distribution} \\  
    \hline
    $\log(M_{\rm BH}/{\rm M}_\odot)$ & $\phantom{7}7.26$ & $-0.14$, $+0.14$ & $-0.42$, $+0.42$\\
    \ml$_{\rm F547M}$ (\Msun/\Lsun) & $\phantom{7}2.19$ & $-0.13$, $+0.13$ & $-0.40$, $+0.39$\\
    \hline
    \multicolumn{4}{l}{\bf AGN light contamination} \\  
    \hline
    $\log(M_{\rm BH}/{\rm M}_\odot)$ & $\phantom{7}6.93$ & $-0.09$, $+0.10$ & $-0.28$, $+0.30$\\
    \ml$_{\rm F547M}$ (\Msun/\Lsun) & $\phantom{7}2.05$ & $-0.07$, $+0.07$ & $-0.21$, $+0.21$\\
    \hline
  \end{tabular}
  \parbox[t]{0.47\textwidth}{\textit{Notes:}{ The search range for $\Gamma$ was set within (0.1--2.0), while the search ranges of $\log(M_{\rm BH}/{\rm M}_\odot)$ and \ml$_{\rm F547M}$ were kept the same as in Table~\ref{fittable}.}}
  \label{tab:bhtest}
\end{table}
%%%%%%%%%%%%%%%%%%%%%%%%%%%%%%%%%%%%%%%%%%%%%%%%%%

%%%%%%%%%%%%%%%%%%%%%%%%%%%%%%%%%%%%%%%%%%%%%%%%
\subsubsection{Axisymmetric ISM distribution}\label{gasdist} 

A different distribution of the molecular gas and dust within the nucleus of NGC~7469 may affect our dynamic results. Following \citet{Davis20}, we tested the impact of the ISM distribution using the axisymmetric assumption by converting the atomic-[CI](1-0) integrated intensity map into the $M_{\rm H_2}$(+ dust) map (I20), then parameterising this ISM-mass map into another MGE form. The dynamical model uses this axisymmetric-ISM-MGE mass (we turned off the {\tt gasGrav} mechanism during the KinMS fit) gives $M_{\rm BH}=1.8^{+3.0}_{-1.1}\times10^7$~\Msun\ and $M/L_{\rm F547M}=2.19^{+0.40}_{-0.39}$ (\Msun/\Lsun), suggesting our derived \Mbh\ is less sensitive to different assumptions of ISM distributions as long as its total mass is determined accurately.

%%%%%%%%%%%%%%%%%%%%%%%%%%%%%%%%%%%%%%%%%%%%%%%%%%
\subsubsection{AGN light contamination}\label{AGN} 

The bright AGN at the centre of NGC~7469 may contaminate a few central pixels of the F547M image. The excess emission here increases the stellar mass density and decreases both the best fits of \Mbh\ and \ml$_{\rm F547M}$ but still within $2\sigma$ CL (see Tables~\ref{fittable} and \ref{tab:bhtest}). To have this conclusion, we re-ran the F547M-photometric-MGE model without masking those central pixels and used it in the KinMS model.  We thus conclude that the AGN-light distribution indeed impacts our dynamical result but not significant.

%%%%%%%%%%%%%%%%%%%%%%%%%%%%%%%%%%%%%%%%%%%%
%%%%%%%%%%%%%%%%%%%%%%%%%%%%%%%%%%%%%%%%%%%%
\section{Discussion}\label{discussion}

%%%%%%%%%%%%%%%%%%%%%%%%%%%%%%%%%%%%%%%%%%
\subsection{Constraining the BLR inclination angle}\label{ffactor}

RM is the commonly use method to reveal SMBHs and constrain their masses in Seyfert 1 AGN using the emission of the broad-lined regions (BLRs) in the time domain rather than resolving it in spatial scale \citep{Blandford82a, Peterson93, Kaspi01, Kaspi05, Bentz06b, Bentz06a, Bentz13}. It uses the intrinsic time variability of AGN emission to measure the time delay between continuum variations in the accretion discs and the broad emission-line response \citep[e.g.][]{Woo02a, Woo02b, McLure04, Shen11, Woo19}. The reverberation time lag for a broad emission line, multiplied by the speed of light $c$, gives a measure of the  response-weighted radius of the BLR. When combined with the velocity width of the broad emission line ($\Delta V$), the \Mbh\ can be estimated using the virial equation:$M_{\rm BH}=\langle f \rangle\times R_{\rm BLR}(\Delta V)^2/G$, where $\langle f \rangle=4.31\pm1.05$ is an average dimensionless scale factor incorporating information of the geometry, kinematics and orientation of the BLR \citep{Grier13a}. However, RM provides a rough-\Mbh\ estimate because (1) the normalisation factor ($\langle f \rangle$) changes significantly in the range of (3.2--7.8) depending on samples and calibration methods \citep{Graham11, Park12, Woo10, Woo13, Onken14} and (2) its uncertainty has a scatter of $>$ 0.5 dex \citep{Onken04} generally unknown for individual objects due to our inability to directly resolve the BLR. These give a concern that non-virial motions in the BLR clouds or radiation pressure might cause large errors in RM-derived $M_{\rm BH}$ \citep{Krolik01}.

The RM-based MBH of NGC~7469 was determined with the canonical normalization factor, and thus there still remains systematic uncertainty given the possibility that the BLR of NGC~7469 may have different geometry and kinematics \citep{Peterson14, Wang14} into account. These particular physical properties could include non-virial velocity components \citep{Denney09, Denney10}, radiation pressure perturbations \citep{Marconi08, Netzer10}, the relative thickness ($h/R_{\rm BLR}$) of the Keplerian BLR orbital plane \citep{Gaskell09}, and the LOS inclination angle ($i$) of this plane \citep{Wills86, Shen14, Runnoe14}. All of these properties were subsumed into the average dimensionless scaling factor for RM AGNs: $\langle f \rangle=4.31\pm1.05$ \citep{Grier13a}. Given our dynamical \Mbh\ constraint of $1.78\times10^7$\Msun\ and its $1\sigma$ uncertainty (Table~\ref{fittable}) and account for the consistency with $1\sigma$ errors of $R_{\rm BLR}$ and $\Delta V$ \citep{Peterson14}, we find $f=7.2^{+4.2}_{-3.4}$ for the BLR in NGC~7469 (replacing $\langle f \rangle$ by $f$). 

We use this specific $f$ to infer the geometry of the BLR based on its analytical expression of a planar BLR of a thick disc, which is given by the expression:$f=\Bigg[4\Bigg(\sin^2i+\Big(\frac{h}{R_{\rm BLR}}\Big)^2\Bigg)\Bigg]^{-1}$, where $i$ is the inclination angle of the system relative to the projected LOS of the Keplerian velocity on the BLR orbital plane \citep{Collin06, Decarli08}. In this scenario, the BLR thickness  may resulted from  radiation pressure of an accretion disc, which creates turbulent motions, disc outflowing winds and non-coplanar orbits \citep{Collin06, Czerny16}. Theoretically, active SMBHs are thought to accrete material in the form of accretion discs and powered by accretion flows. The accretion discs convert gravitational energy into intense radiation \citep{Shakura73}, making gas in the BLR move with velocities of thousands of kilometers per second. Under virial equilibrium, the \Mbh\ can be determined from the intrinsic FWHM$_{\rm int}$ of observed lines by using the following equation: $M_{\rm BH}=fG^{-1}R_{\rm BLR}{\rm FWHM_{int}^2}$. In practice, however, the $f$ factor derived by comparing single epoch \Mbh\ with that masses obtained from the \Mbh--$L$ \citep{Decarli08} or \Mbh--$\sigma$ \citep{Shen14} scaling relations of spheroidal galaxies, or from amplitude of the excess X-ray variability variance \citep{Nikolajuk06} reveal the the anti-correlation between $f$ and ${\rm FWHM_{obs}}$ of the broad emission lines ($f\propto {\rm FWHM_{obs}^{-1}}$), implying that ${\rm FWHM_{int}\propto FWHM_{obs}^{1/2}}$.

Recently, \citet{Mejia-Restrepo18} used a sample of 39 $z\sim1.55$, high signal-to-noise spectroscopic type-1 AGN observed with VLT/X-Shooter spectrograph to perform Monte Carlo simulations of the LOS inclination dependence of FWHM$_{\rm int}$ of the virialised velocity component of the BLR. They found that only thin BLRs ($h/R_{\rm BLR}\lesssim0.1$) can reproduce the observed bi-dimensional distribution of $f\propto {\rm FWHM_{obs}^{-1}}$ and the predicted FWHM$_{\rm int}$ $\propto$ FWHM$_{\rm obs}^{1/2}$. We thus consider the thin disc BLR model for the NGC~7469 AGN in which $(h/R_{\rm BLR})\approx0$, giving the BLR's LOS inclination of $i\approx11.0^\circ$$_{-2.5}^{+2.2}$, which tightly constrains the face-on orientation of the BLR. However, the physics of a compact BLR is likely to remain constant over time (i.e. consistent in terms of the Eddington ratio), making it is possible to study active SMBHs at low and high redshifts \citep{Mejia-Restrepo18}.

%%%%%%%%%%%%%%%%%%%%%%%%%%%%%%%%%%%%%%%%%%%%
\subsection{A new game changer for constraining \Mbh\ in AGN}\label{newgame}

%%%%%%%%%%%%%%%%%%%%%%%%%%%%%%%%%%%%%%%%%
\begin{figure*}  
	\includegraphics[scale=0.08]{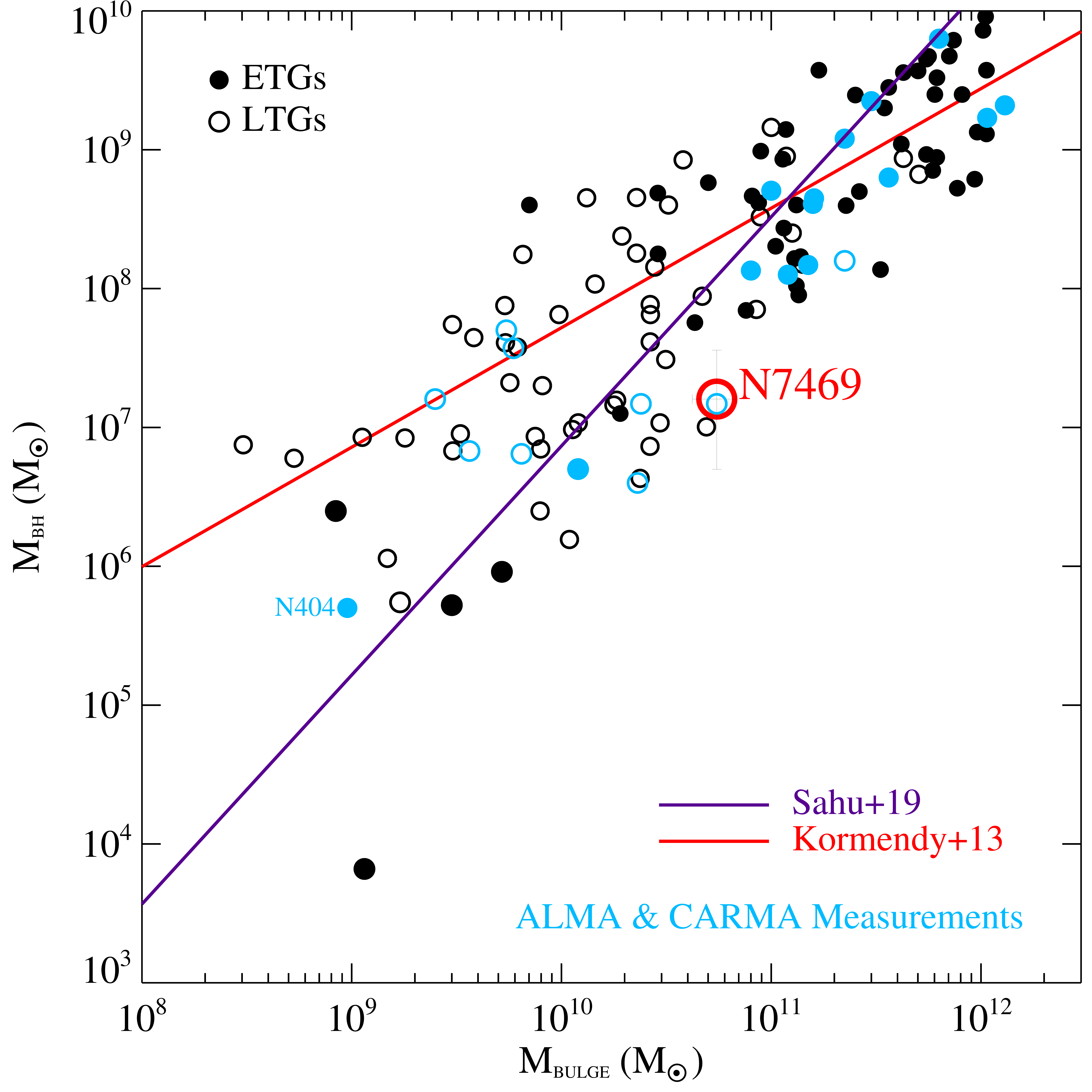} \hspace{5mm}
	\includegraphics[scale=0.08]{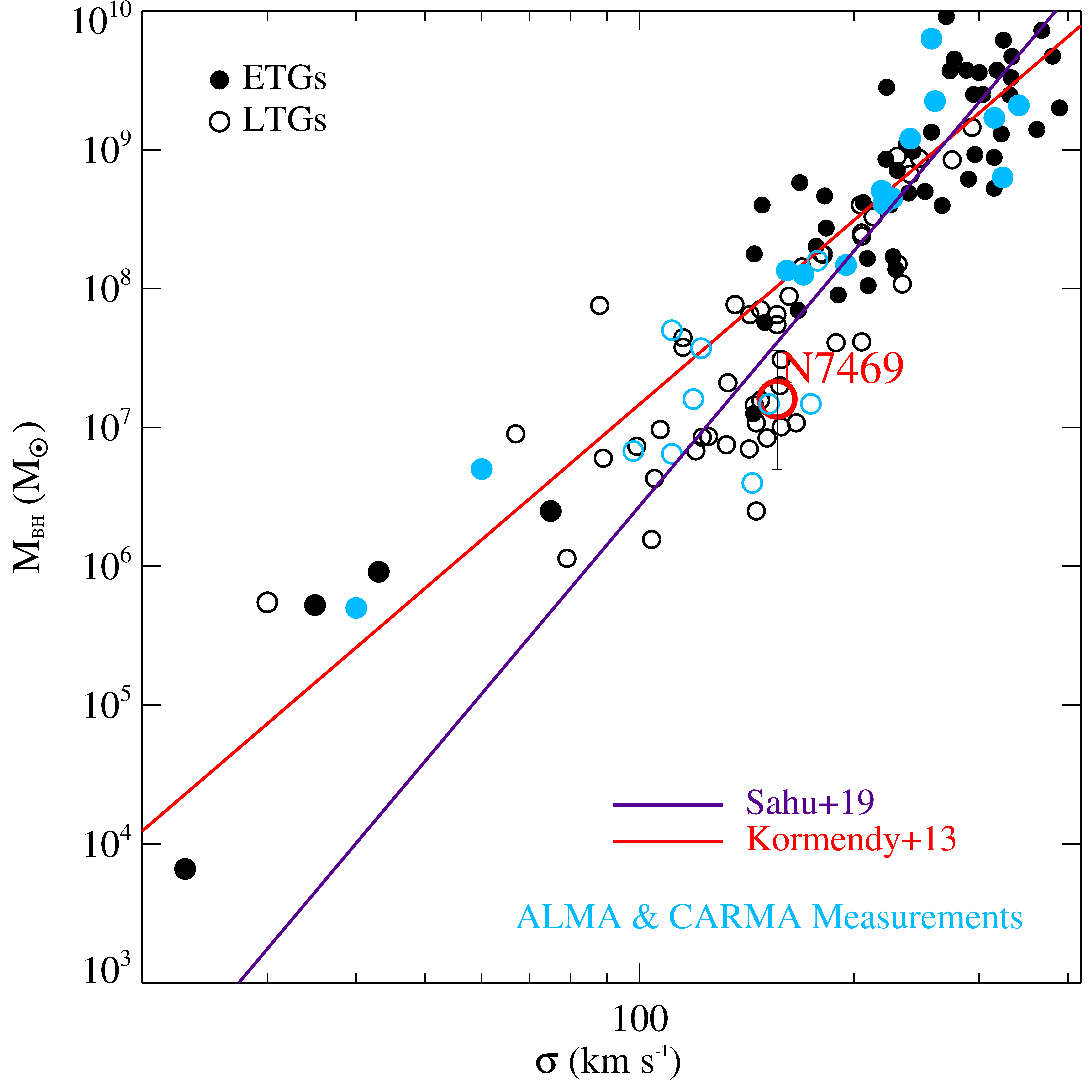}
   	\caption{Our NGC~7469 \Mbh~(red encircle) in the context of the \Mbh--$M_{\rm bulge}$ (left) and \Mbh--$\sigma_\star$ scaling relations (right). 24 molecular-gas-\Mbh\ measurements using both ALMA \citep[][Nguyen et al. submitted, Smith et al. submitted]{Davis17, Davis18, Davis20, Onishi15, Barth16a, Barth16b, Boizelle19, Boizelle20, Combes19, Smith19, Nagai19, North19, Thater19b, Nguyen20} and CARMA \citep{Davis13, Onishi17} observations are plotted in cyan. The scaling relations of \citet{Sahu19a} and \citet{Kormendy13} are plotted in the specific-solid-colour lines shown in the legend. \Mbulge\ are taken from \citet{McConnell13}, \citet{Krajnovic13}, \citet{Salo15}, \citet{Huang16}, \citet{Savorgnan16}, and \citet{Sani18}, while $\sigma_\star$ are taken from \citet{Ho09}, \citet{Seth10a}, \citet{Coccato13}, \citet{Krajnovic13}, \citet{Kormendy13}, \citet{Savorgnan16}, and \citet{Sani18}. Error bars of our measurement is 3$\sigma$ uncertainty.}
    \label{scalingplot1}   
\end{figure*}
%%%%%%%%%%%%%%%%%%%%%%%%%%%%%%%%%%%%%%%%%

This work shows that the \Mbh\ in the AGN type 1 NGC~7469 can be constrained using the atomic-[CI](1-0) kinematics. The atomic-[CI](1-0) emission line is bright, strongly peaked toward the galaxy centre at the CND-scale due to the XDR around the AGN (I20). These unique features suggest that the atomic-[CI](1-0) line must be more appropriate to measure central \Mbh\ in active galaxies than the low-J CO-molecular lines. This newly precise measurement of \Mbh\ using ALMA atomic-gas dynamics can also be used to constrain the virial geometric factor $f$ of the BLRs in Seyfert 1 AGN NGC~7469, then provide the first insights into the unresolved structure of BLRs \citep[except the BLR of 3C 273 seems to be spatially resolved by using GRAVITY;][]{GravityCollaboration18} as the effect of line-of-sight (LOS) inclination ($i$) in a planar distribution of the BLR emitting gas in which $f(i)$ seems following an anti-correlation \citep{Mejia-Restrepo18}. 

Constraining a functional form for such an important correlation based on a large sample of nearby AGN will provide a better understanding on the correlation between the \Mbh\ and AGN properties (e.g. FWHM and the velocity dispersion of the emitting gas in the BLR) that will help to derive more accurate RM--based \Mbh\ in more distant AGN beyond the local Universe ($z\gtrsim0.1$). Besides, these precise measurements of \Mbh\ will ultimately be a key to understand the \Mbh\ function and the mean radiative efficiency of SMBH accretion \citep{Shankar09, Shankar16}. Fortunately, high-spatial-resolution and high sensitivity of ALMA observations now allow us to probe the X-ray dissociated atomic-gas kinematics close to the SMBHs' SOI and infer their masses accurately through the resolved manner (i.e. it can achieve the synthesised beam of $\theta_{\rm [CI](1-0)}\approx0\farcs009$ (or 9 mas) with the most extended basedline of 16~km), which was impossible in previous studies (i.e. both RM and dynamics with low-J CO molecular tracers). This work pioneers the usage of the atomic-[CI](1-0) line as a new kinematics tracer and can be a game-changer in the \Mbh\ measurement field, at least for the active galaxies where other methods cannot work well. More importantly, the atomic-[CI](1-0) emission has a higher rest-frequency than low-J CO-molecular lines, suitable for performing similar dynamical observations/measurements toward higher redshift objects (i.e. the synthesis beam of $\theta_{\rm [CI](1-0)}\approx9$ mas is good enough for constraining any black holes with \Mbh~$>9\times10^8$~\Msun\ at all redshift\footnote{Note that the curve $r_{\rm SOI}$ varies as a function of both redshift ($z$) and black hole mass (\Mbh), $r_{\rm SOI}\propto f(z, M_{\rm BH})$, which has the minimum at $z\approx1$. Given the current majority measurements in the local Universe with ALMA observations that have $\theta_{\rm beam}>2\times r_{\rm SOI}$ (references in the text), the speculated \Mbh\ values could be smaller by a factor of a few. For example, \citet{Nguyen20} constrain the \Mbh\ in NGC~3504 at the observational scale of six times larger than its $r_{\rm SOI}$. We thus argue that depending on the case, \Mbh\ could be lower by an appropriate factor in comparison to the value inferred from the function $r_{\rm SOI}\propto f(z, M_{\rm BH})$.}). Thus, we can extend this atomic-[CI](1-0) line survey and analysis for high redshift AGN in the future. 

%%%%%%%%%%%%%%%%%%%%%%%%%%%%%%%%%%%%%%%%%%%%%
\subsection{Comparison to RM-based \Mbh\ and scaling relations}\label{scaling}

As discussed in Section \ref{sec:intro}, the central \Mbh\ of NGC~7469 was determined via both the direct signatures of radio and X-ray radiation on the fundamental plane and the indirect signature of the BLR variability. However, these \Mbh\ occupy a wide range of mass from $10^5$ to $10^8$\Msun. Our gas-dynamical models using ALMA observation in this work give a $\approx40$\% higher value for the \Mbh\ than the estimate of \citet{Peterson14} and \citet{Wang14} using RM of the H$\beta$ and \ion{He}{ii} $\lambda$4686 emission lines, respectively.  

Our derived \Mbh\ of NGC~7469 is consistent with the empirical \Mbh--\Mbulge\ correlation of both ETGs and late-type galaxies \citep[LTGs;][]{Sahu19a, Sahu19b} within 3$\sigma$ uncertainty. However, it is about one order of magnitude below the same correlations compiled mainly for higher mass galaxies that have S\'ersic profiles without central cores of \citet{Kormendy13}, \citet{McConnell13}, and \citet{Saglia16}, as well as 1.5 magnitudes below the same correlation of \citet{Scott13}, compiled for the core-S\'ersic and lower-mass galaxies ($M_\star\lesssim2\times10^{10}$~\Msun). It is also well below the \citet{Greene20} correlation with the same amount of magnitude compiled for all types and masses of galaxies. In the context of \Mbh--$\sigma_\star$, the NGC~7469 \Mbh\ is consistent with the \citet{McConnell13}, \citet{Kormendy13}, \citet{Saglia16} and \citet{Sahu19a, Sahu19b} correlations rather than the correlation of \citet{Greene20}. We show our measurement of NGC~7469 in the contexts of the \citet{Kormendy13} and \citet{Sahu19a} \Mbh--\Mbulge\ and \Mbh--$\sigma_\star$ correlations in Fig.~\ref{scalingplot1} only. Here, the former is the most high-cited scaling relation, while the latter is the recent most completed compilation correlation with adding new measurements and revisions from 2013.

\citet{Smith20} recently investigated the empirical correlations between \Mbh\ and the flat rotation velocities measured either at a large radius in their rotation curves or via spatially-integrated emission line widths of a sample of both spatially resolved and unresolved galaxies (so-called $\Delta V_{\rm CO}$--\Mbh\ correlation), that have \Mbh\ constrained via molecular-gas-dynamical modellings. Their spatially resolved sub-sample includes 27 galaxies with CO-CND detected, and nine of them have dynamical-\Mbh\ constrained via ALMA in the same manner we did here for NGC~7469. The unresolved sub-sample, on the other hand, includes 24 same targets with six ALMA dynamical-\Mbh. Here, they assumed the rotation velocity traced by the de-projected integrated CO emission line width and found a tight correlation of 24 spatially resolved CO discs (three targets were omitted): $\log(M_{\rm BH}/$M$_{\odot}$) = $(7.5\pm0.1) + (8.5\pm0.9)[\log(W_{50}/\sin i\ {\rm km\;s}^{-1}) - 2.7]$, where $W_{50}$ is the FWHM of a double-horned emission line profile and $i$ the inclination of the CO disc. Another tight correlation between this de-projected CO line widths (flat rotation velocity) and the stellar-velocity dispersion averaged within one effective radius of $\log(\sigma_\star/{\rm km\;s}^{-1}) = (2.20\pm0.02) + (1.1\pm0.1)[\log(W_{50}/\sin i\ {\rm km\;s}^{-1}) - 2.7]$ was also found, which is so-called the $\Delta V_{\rm CO}$--$\sigma_\star$ correlation. In the context of the $\Delta V_{\rm CO}$--\Mbh\ correlation and with $\Delta V_{\rm CO}\approx400$~\kms\ (Section~\ref{almaobs}), NGC~7469 is about $\approx0.2$ dex of out $1\sigma$ intrinsic scatter (0.5 dex). This outlier may be caused by the nearly face-on orientation of the CO/[CI] disc, implying that the inclination uncertainties are very large. However, in terms of the $\Delta V_{\rm CO}$--$\sigma_\star$ correlation, the \Mbh\ of NGC~7469 is well predicted (scatter $\lesssim0.1$ dex). As companions to other \Mbh--galaxy's properties scaling relations, these two correlations can also be used to estimate the local-\Mbh\ function.

%%%%%%%%%%%%%%%%%%%%%%%%%%%%%%%%%%%%%%%%%%%%
%%%%%%%%%%%%%%%%%%%%%%%%%%%%%%%%%%%%%%%%%%%%
\section{Conclusions}\label{conclusions}

We present dynamical-mass measurements for the central SMBH in the AGN type 1 NGC~7469 using \hst\ imaging and the bright atomic-[CI](1-0) and molecular-$^{12}$CO(1-0) emissions observed by ALMA.  We highlight our main results below:

\begin{enumerate}
	 \item NGC~7469 hosts multiple transitions of molecular-CO and atomic-[CI] lines within the CND and starburst ring. Detailed analysis of the $^{12}$CO(1-0) and [CI](1-0) emissions find a dynamically settled CND in the inner $1\arcsec$ of the galaxy centre.
	 
	 \item Our dynamical modelling for the atomic-[CI](1-0) data suggests the presence of a central SMBH of $M_{\rm BH}=1.78^{+2.69}_{-1.10}\times10^7$~\Msun, while that of the molecular-$^{12}$CO(1-0) data provides a mass of $M_{\rm BH}=1.60^{+11.52}_{-1.45}\times10^7$~\Msun, a factor of $\approx(1.5$--$2)$ higher than the \citet{Peterson14} RM-based-\Mbh. These two models also find a consistent $M/L_{\rm F547M}\approx2.20$ (\Msun/\Lsun). 
	 
	 \item The atomic-[CI](1-0) emission may be the best line for performing the gas-dynamical modelling and constraining central \Mbh\ in active galaxies as [CI](1-0) morphological distribution is closer to the SMBH and brighter than the low-J CO lines. Thus, it is an excellent tracer to do cross-checks between RM-based~and gas-dynamical methods in near and far AGN. However, we should caution that users need to test this point further with other AGN and higher spatial resolution data.
	 
	  \item Our new \Mbh\ estimate for NGC~7469 provides a specific value for the RM AGN dimensionless scaling factor of the BLR of $f=7.2^{+4.2}_{-3.4}$ for the first time. This value is $\approx$ 40\% higher than the average value of $\langle f \rangle$ applied for all previous indirect RM-based \Mbh\ estimates in Seyfert 1 AGN. It also reveals a thin accretion disc for the unresolved BLR, which is oriented face-on with the LOS inclination of $i\approx11.0^\circ$$_{-2.5}^{+2.2}$.  
	 
	 \item Our \Mbh\ is consistent with the empirical \Mbh--\Mbulge\ correlations of the recent compilation of \citet{Sahu19b} for all types of galaxies, which have direct-\Mbh\ measurements. However, it is almost one order of magnitude below the correlations of \citet{Kormendy13}, \citet{McConnell13}, and \citet{Saglia16} of the more massive SMBHs and galaxies without the S\'ersic bulges in their surface brightness profiles. This negative offset is even larger (roughly 1.5 magnitudes) in the context of the \citet{Greene20} scaling relation compiled for all types and masses of galaxies and the \citet{Scott13} correlation compiled for lower-mass galaxies with S\'ersic cores. In the context of the \Mbh--$\sigma_\star$ correlation, the \Mbh\ seems consistent with the \citet{Kormendy13}, \citet{McConnell13}, \citet{Saglia16}, and \citet{Sahu19a, Sahu19b} scaling relations but inconsistent with the compilation of \citet{Greene20}, which is above the locations of the SMBH about one order of magnitude.  Our \Mbh\ is also consistent with the $\Delta V_{\rm CO}$--$\sigma_\star$ correlation but offset 0.2 dex outside $1\sigma$ (0.5 dex) scatter of the $\Delta V_{\rm CO}$--\Mbh\ scaling relation, which are recently compiled by \citet{Smith20}.
		 
	 \item The NGC~7469's \Mbh\ has been constrained consistently using molecular/atomic gas of ALMA, although the observational scales of the data is a factor of $\approx30$ times larger than the angle subtended by the SMBH's $r_{\rm SOI}$. Because of the poorly resolvable scale of our data, we argue this consistency applies to the case of NGC~7469 only. 	 	 
	 
	 \item This work pioneers the user of the atomic-[CI] line in constraining \Mbh\ in AGN accurately, which then allows estimating the BLR's inclination angle, a crucial ingredient to understand the unified AGN model.
\end{enumerate}

%%%%%%%%%%%%%%%%%%%%%%%%%%%%%%%%%%%%%%%
%%%%%%%%%%%%%%%%%%%%%%%%%%%%%%%%%%%%%%%
\section*{ACKNOWLEDGEMENTS}

%%%%%%%%%%%%%%%%%%%%%%%%%%%%%%%%%%%%%%%%%%%%%%%%%%
The authors would like to thank the anonymous referee for their careful reading and useful comments, that helped to improve the paper greatly. The authors also thank National Astronomical Observatory of Japan (NAOJ), National Institute of Natural Sciences (NINS), and the PHENIKAA Institute for Advanced Study (PIAS of PHENIKAA University for supporting this work. T.I., S.B., and K.K. gratitude to the support from the Japan Society for the Promotion of Science (JSPS) KAKENHI grant \#JP20K14531, 19J00892 and JP17H06130 (and the NAOJ ALMA Scientific Research Grant Number 2017-06B), respectively. S.T. acknowledges funding from the European Research Council (ERC) under the European Union's Horizon 2020 research and innovation programme under grant agreement No 724857 (Consolidator Grant ArcheoDyn). 

This paper makes use of the following ALMA data: ADS/JAO.ALMA \#2017.1.00078.S. ALMA is a partnership of ESO (representing its member states), NSF (USA) and NINS (Japan), together with NRC (Canada) and NSC and ASIAA (Taiwan) and KASI (Republic of Korea), in cooperation with the Republic of Chile. The Joint ALMA Observatory is operated by ESO, AUI/NRAO and NAOJ. We also thank the ALMA operators and staff and the ALMA help desk for diligent feedback and invaluable assistance in processing these data.

{\it Facilities:} ALMA and \hst/UVIS~WFC3.

{\it Software:} \texttt{IDL}, \texttt{CASA}, \texttt{astropy}, \texttt{emcee}, \texttt{KinMS}, 
\texttt{Kinemetry}, and \texttt{MgeFit}.

%%%%%%%%%%%%%%%%%%%%%%%%%%%%%%%%%%%%%%%
%%%%%%%%%%%%%%%%%%%%%%%%%%%%%%%%%%%%%%%
\section*{Data Avaibility}

The data underlying this article will be shared on reasonable request to the corresponding author.

\bibliographystyle{mnras}
\bibliography{NGC7469BH_v4} % if your bibtex file is called example.bib

%%%%%%%%%%%%%%%%%%%%%%%%%%%%%%%%%%%%%%%%%%%%%%%%%%
% Don't change these lines
\bsp	% typesetting comment
\label{lastpage}
\end{document}